# Towards hydrogen-rich ionic (NH$_4$)(BH$_3$NH$_2$BH$_2$NH$_2$BH$_3$) and related molecular NH$_3$BH$_2$NH$_2$BH$_2$NH$_2$BH$_3$[#]

Rafał Owarzany*[a], Tomasz Jaroń[b], Krzysztof Kazimierczuk[a], Przemysław J. Malinowski[a], Wojciech Grochala[a] and Karol J. Fijalkowski*[a]

Attempts of synthesis of ionic (NH$_4$)(BH$_3$NH$_2$BH$_2$NH$_2$BH$_3$) using metathetical approach resulted in a mixture of the target compound and a partly dehydrogenated molecular NH$_3$BH$_2$NH$_2$BH$_2$NH$_2$BH$_3$ product. The mixed specimen was characterized by NMR and vibrational spectroscopies, and the crystal structure of their cocrystal was solved from powder x-ray diffraction data, and supplemented by theoretical density functional theory calculations. Despite their impressive hydrogen content, and similarly to ammonia borane, both title compounds release hydrogen substantially polluted with borazine, and traces of ammonia and diborane.

## Introduction

Protic-hydridic compounds constitute one important family of solid-state hydrogen storage materials with the potential to be applied as onboard fuel systems required in the hydrogen economy. The presence of both positively and negatively charged hydrogen atoms results in the formation of a network of dihydrogen bonds governing the crystal structure and facilitating the process of thermal decomposition.[1] These features were observed and thoroughly described for ammonia borane,[2] metal amidoboranes,[3] and further explored for NH$_4$BH$_4$.[4]

Ammonia borane (AB) is one of the best-researched materials in this group, being an air and water insensitive solid and containing *ca.* 19.6% of hydrogen by weight.[5] Unfortunately, AB releases only 1/3 of the stored hydrogen below 120°C.[6] Moreover, the hydrogen released is contaminated with ammonia, diborane, borazine, aminoborane and aminodiborane, which excludes its use as a direct H$_2$ source for low-temperature fuel cells.[7] Such high gravimetric H content, however, provides significant room for modifications – even if the relatively heavy elements are introduced, the system still should be able to fulfil the gravimetric DOE requirements for H storage materials (Fig. 1).[8]

Among the derivatives of AB, amidoborane salts of a general formula M(NH$_2$BH$_3$)$_n$ [abbreviated here as MAB or M(AB)$_n$] constitute the largest group.[9,10] Two dozens of mono- and bimetallic amidoborane salts have been reported. Some of them [*i.e.* KAB,[11] RbAB,[12,13] CsAB,[12,13] Mg(AB)$_2$,[14] Ba(AB)$_2$,[15] Al(AB)$_3$,[16] LiAl(AB)$_4$,[17] Li$_2$Mg(AB)$_4$[18], ] evolve pure H$_2$ upon thermal decomposition at *ca.* 100°C. Nonetheless, all these materials suffer from a lack of reversibility and relatively low hydrogen content available at moderate temperatures.[9,10]

Recently, a novel group of ammonia borane derivatives containing five-membered chain anions of a general formula M(BH$_3$NH$_2$BH$_2$NH$_2$BH$_3$) [abbreviated here as M(B3N2)] have been reported.[19–23] Among them, one can list two allotropes of Verkade's base salt[20,21], five alkali metal salts[19,21,22,24,25] and four ionic liquids.[23]. Although three of them [*i.e.* Li(B3N2)[21,22], (Bu$_4$N)(B3N2),[23] (Et$_4$N)(B3N2)[23]] meet the target H wt% content and release pure hydrogen below 150°C, yet none of them fulfils all the DOE targets simultaneously.[8]

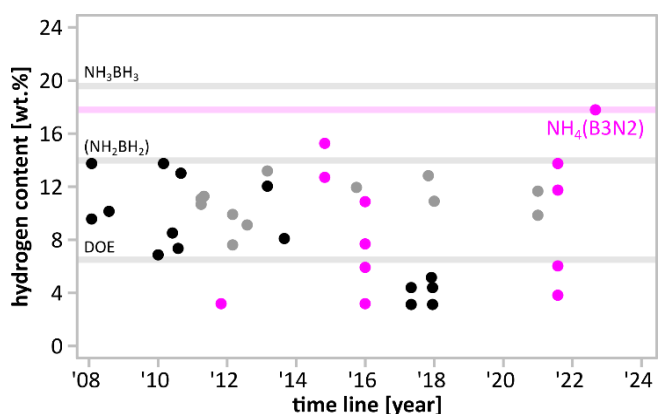

**Fig. 1.** Hydrogen content of monometallic amidoboranes (black), bimetallic amidoboranes (grey), M(B3N2) salts (magenta) as a function of reporting date. Hydrogen content of NH$_3$BH$_3$ (19.6%) polymeric (NH$_2$BH$_2$) (14.0%) and DOE ultimate target (6.5%) given as a reference.

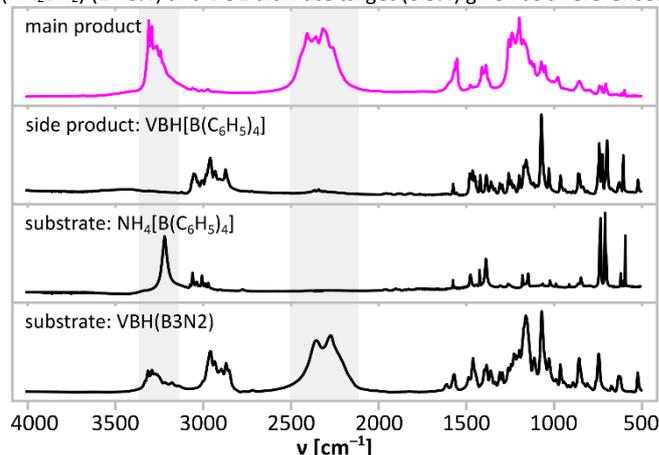

**Fig. 2.** Comparison of IR absorption spectra of precursors and products of metathetic synthesis performed according to Eq. 1. NH stretching and BH stretching regions are marked with grey fields.

# Results and discussion

## 2.1 Synthesis

Synthesis of (NH$_4$)(B3N2) was attempted employing Jaron's *et al.* metathetic approach mediated by the precursors containing weakly coordinating ions [26–28]. The reaction was conducted in dry THF similarly to the previous syntheses of all alkali metal M(B3N2) salts[21] according to Equation 1:

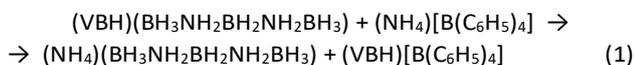

(VBH)(BH$_3$NH$_2$BH$_2$NH$_2$BH$_3$) + (NH$_4$)[B(C$_6$H$_5$)$_4$] →
→ (NH$_4$)(BH$_3$NH$_2$BH$_2$NH$_2$BH$_3$) + (VBH)[B(C$_6$H$_5$)$_4$]   (1)

Unexpectedly, during the reaction, we observed the evolution of a small amount of gas which should not occur in a metathetic reaction. Since the expected main product contains ammonium cation and B3N2$^-$ anion (essentially, a derivative of a borohydride anion) we assumed that – similarly to what is observed for NH$_4$BH$_4$[29] – hydrogen might be evolved upon reaction of these ions according to Equation 2:

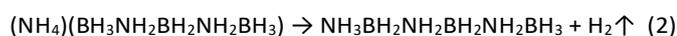

(NH$_4$)(BH$_3$NH$_2$BH$_2$NH$_2$BH$_3$) → NH$_3$BH$_2$NH$_2$BH$_2$NH$_2$BH$_3$ + H$_2$↑   (2)

Confirmation that this surmise is correct is presented below.

Synthesis led to a mixture of well THF-soluble products, which were separated by precipitation of the side product (VBH)[B(C$_6$H$_5$)$_4$] in dry DCM. Both products were subjected to spectroscopic analyses (Fig. 2) and powder X-ray diffraction (Fig. 3) to demonstrate successful ion exchange according to Eq. 1. Indeed, FTIR spectra (Fig. 2) of the products show that the target product contains the NH and BH groups, while the side product contains the CH groups. Unfortunately, complete separation of the main product and side product was not achieved, as documented by very weak CH bands at *ca.* 3000 cm$^{-1}$ from B(C$_6$H$_5$)$_4^-$ anions for the former, and very weak BH bands at *ca.* 2400 cm$^{-1}$ from (B3N2)$^-$ anions for the latter product. X-ray diffraction points to the same conclusion, showing that two new distinct crystalline species formed during the reaction (Fig. 3). The diffraction patterns of the products are free from reflections coming from the substrates which suggests at least 95% purity of the former.

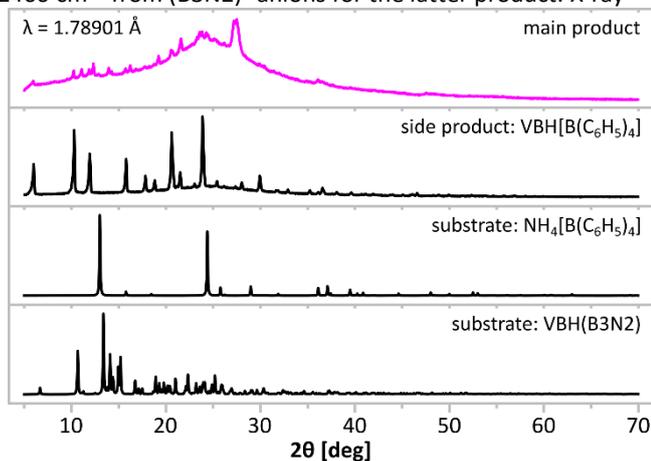

**Fig. 3.** Comparison of powder X-ray diffraction patterns of precursors and products of metathetic synthesis performed according to Eq. 1. CoK$_{\alpha 1,2}$, λ = 1.78901 Å.

## 2.2 NMR spectra

A detailed $^{11}$B NMR investigation of the main product dissolved in THF-d$_8$ was performed (Fig. 4). A typical spectrum of M(B3N2) salt consists of a triplet at *ca.* −8.5 ppm from [BH2] groups and a quartet at *ca.* −22.5 ppm from [BH3] groups, having a relative intensity of 1:2).[20,21] Here, the spectrum of the main product is more complicated and contains two triplets (δ = −10.4 ppm, J = 101 Hz; δ = −12.3 ppm, J = 102 Hz) and a quartet (δ = −22.2 ppm, J = 91 Hz) in an intensity ratio that varies from batch to batch (in average *ca.* 4:3:5). These features altogether suggest that the main product formed according to (Eq. 1) partially undergoes a subsequent dehydrogenation reaction (Eq. 2). Variation in the observed intensity ratio of the signals may be caused by the partial decomposition of the main product thus changing the ratio between the components of the product. We note that a 1:3 mixture of (NH$_4$)(B3N2) and NH$_3$BH$_2$NH$_2$BH$_2$NH$_2$BH$_3$ (abbreviated as N3B3) would yield 4 BH$_2$ triplets units coordinated by two NH$_2$ groups, 3 BH$_2$ triplets from units coordinated by NH$_2$ and NH$_3$, and 5 BH$_3$ quartets altogether. This result would suggest that ca. ¾ of (NH$_4$)(B3N2) decomposed to N3B3 while dissolved in THF.

To get more insights into the processes occurring during the synthesis, we conducted *in situ* $^{11}$B{$^1$H} NMR measurements in THF-d$_8$ to monitor signals of the products and substrates (Fig. 5). The monitoring showed that all three signals from the product(s) (marked with #) arise simultaneously, testifying to the simultaneous progress of reactions described by Eq.1 and Eq.2. Apart from the signals assigned to the substrates and the main product, numerous additional signals which do not change during the reaction are present. These signals come from the moieties that do not play a direct role in the formation of the main product, *e.g.* [B(C$_6$H$_5$)$_4$]$^-$, in which the chemical neighbourhood of boron atom does not change during the synthesis. The monitored reaction (Fig. 4) was not completed (i.e. signals of the substrates were still intense) because of the local depletion of the substrates (no mixing was applied in an NMR test tube inside the spectrometer).

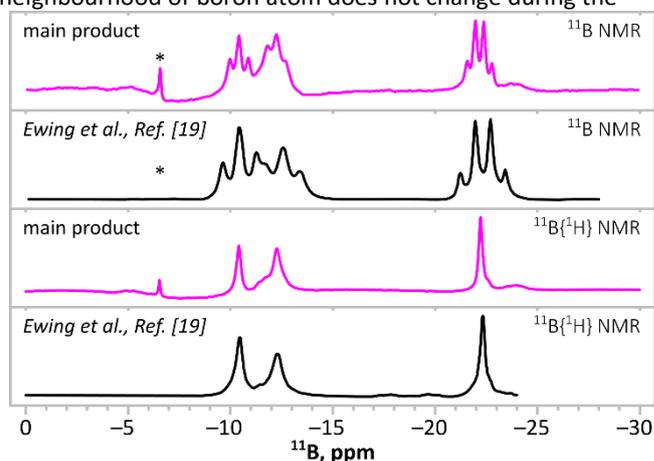

**Fig. 4.** Comparison of $^{11}$B NMR spectra of the main product of the synthesis according to Eq. 1 (magenta lines) with the spectra of a product of reaction according to Eq. 2 reported by Ewing et al.[19] (black lines) both with (bottom spectrum). Boron spectra show with and without $^1$H decoupling. * indicates [B(C$_6$H$_5$)$_4$]$^-$ anions.

It is worth mentioning that $^{11}$B NMR spectrum of the main product(s) is very similar to the spectrum reported by Ewing et al. in 2013, having the same pattern of three signals: two triplets (δ = −10.5 ppm, J = 95 Hz; δ = −12.4 ppm, J = 104 Hz) and a quartet (δ = −22.3 ppm, J = 95 Hz)[19] with 4.1:3.2:5.0 intensity ratio of the signals (according to our analysis of graphical data show in that study, Fig. 9a in Ref. [19]). The report of Ewing et al. focused on the synthesis of a neutral 6-membered chain molecule, B3N3, via a direct reaction between Na(B3N2) and ammonium chloride, according to the following equation:[19]

$$Na(BH_3NH_2BH_2NH_2BH_3) + NH_4Cl \rightarrow$$
$$\rightarrow NH_3BH_2NH_2BH_2NH_2BH_3 + H_2\uparrow + NaCl\downarrow \quad (3)$$

The reaction, performed in glyme solution, was accompanied by evolution of hydrogen gas, which is similar to our observations.

Ewing et al. assumed that the solid product obtained was B3N3 based only on an NMR study of this material.[19] Three signals observed were assigned to three boron-containing groups present in the molecule (two BH$_2$, one BH$_3$). In the case of successful synthesis of B3N3, however, all signals observed should be equally intense (1:1:1) while their intensity ratio was clearly different (Fig. 4). Our analysis suggest that the reaction towards B3N3 is in fact a two-step process (Eqn. 1 and 2) and some (NH$_4$)(B3N2) intermediate (i.e. our main target compound) remains. The resulting assignment of NMR signals for both products is given in Table 1.

To further support our claim, we performed further characterisation of the reaction product.

## 2.3 FTIR and Raman analysis

Two preeminent sets of bands in the vibrational (IR absorption and Raman scattering) spectra of the main product (Fig. 6, Fig. 7), originate from stretching vibrations of NH (3000–3400 cm$^{-1}$) and BH (2150–2400 cm$^{-1}$) groups. They are accompanied by bands coming from deformation vibrations of NH$_x$ moieties (1400–1600 cm$^{-1}$) as typical of M(B3N2) salts and by BN stretching. BH$_x$ deformation modes fall below 1350 cm$^{-1}$.

Aside from the bands typical for M(B3N2) salts, the NH stretching region of Raman spectrum (Fig. 7) contains a relatively low frequency band (peaking at 3041 cm$^{-1}$) originating from ammonium cations. The ammonium cations are known to give strong Raman at much lower energies than the [NH$_2$] and [NH$_3$] groups, e.g. ammonium chloride gives a single band at 3052 cm$^{-1}$,[30] while ammonium borohydride yields two bands at somewhat higher energies of 3118 cm$^{-1}$ and 3178 cm$^{-1}$.[4,31]

In the higher energy part of NH stretching region, in both FTIR and Raman spectra, we observe at least 6 distinct bands, four of which (3306 cm$^{-1}$, 3288 cm$^{-1}$, 3259 cm$^{-1}$, 3239 cm$^{-1}$ in IR; 3307 cm$^{-1}$, 3288 cm$^{-1}$, 3260 cm$^{-1}$, 3240 cm$^{-1}$ in Raman) form a doublet of doublets seen for heavy alkali metal M(B3N2) salts; it is characteristic for M(B3N2) salts that that the higher energy doublet is more intense in FTIR spectra, while the lower-wavenumber one is stronger in Raman spectra. In contrast, the lower energy one is more intensive in Raman spectra.[21] The presence of such doublets is caused by Davydov splitting,[32] i.e. the interaction of [NH$_2$] groups coexisting within one crystallographic unit cell (the resonance of the corresponding oscillators removing their degeneration). A similar split of NH band was observed in the spectra of Rb(B3N2) and Cs(B3N2), which contain gauche-form of (B3N2)$^-$ anions, unlike the lighter analogues featuring straight anions and not showing Davidov split.[21] The split observed here equals ±9 cm$^{-1}$, which is intermediate between those of ±4 cm$^{-1}$ and ±14 cm$^{-1}$ seen for Rb(B3N2) and Cs(B3N2), respectively.[21]

**Table 1.** Chemical shifts, J-coupling values and assignment of the signals observed in $^{11}$B NMR spectra of the main product, sample reported by Ewing et al.[19]. Data for alkali metal M(B3N2) salts[21] given as reference.

| compound | −NH$_2$−**BH$_2$**−NH$_2$− | | NH$_3$−**BH$_2$**−NH$_2$− | | **BH$_3$**−NH$_2$− | |
|---|---|---|---|---|---|---|
| | δ [ppm] | J [Hz] | δ [ppm] | J [Hz] | δ [ppm] | J [Hz] |
| main product | −10.4 | 101 | −12.3 | 102 | −22.2 | 91 |
| Ewing et al.[19] | −10.5 | 95 | −12.4 | 104 | −22.3 | 95 |
| (VBH)(B3N2)[21] | −8.2 | 100 | − | − | −21.6 | 91 |
| Li(B3N2)[21] | −8.4 | 103 | − | − | −22.6 | 90 |
| Na(B3N2)[21] | −8.7 | 99 | − | − | −22.4 | 91 |
| K(B3N2)[21] | −8.6 | 101 | − | − | −22.0 | 89 |
| Rb(B3N2)[21] | −8.4 | 100 | − | − | −21.7 | 90 |
| Cs(B3N2)[21] | −8.4 | 101 | − | − | −21.2 | 94 |

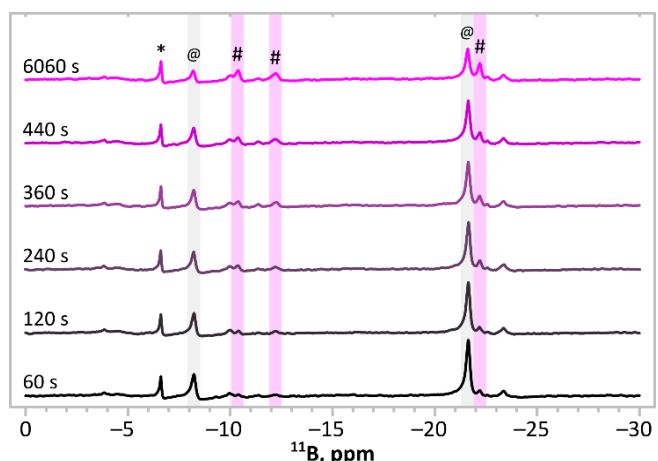

**Fig. 5.** The sequence of $^{11}$B{$^1$H} NMR spectra collected in situ upon synthesis according to Eq. 1. The bottom spectrum shows the mixture of substrates (t=60s). The top spectrum shows to final mixture of products and substrates (t=6060s). @ indicates signals from substrates. # indicates signals of the main product. * indicates [B(C$_6$H$_5$)$_4$]$^-$ anions.

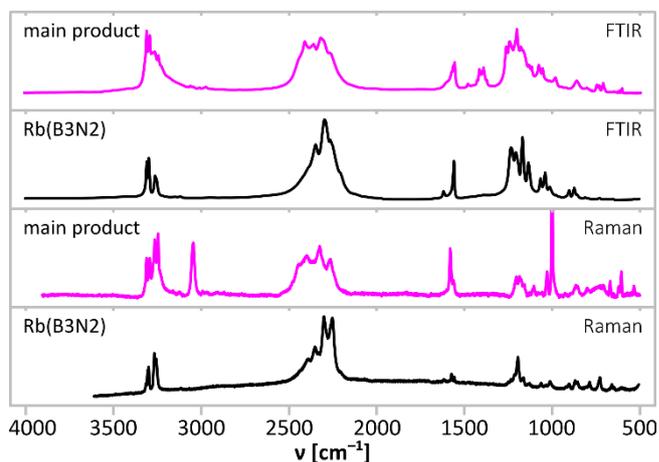

**Fig. 6.** Comparison of FTIR absorption (top in each bracket) and Raman scattering (bottom in each bracket) spectra of the main product of the synthesis according to Eq. 1 (magenta) and the spectra of alkali metal M(B3N2) salts. Regions magnified in Fig. 7 (NH stretching and NH bending) are marked with grey fields.

Two remaining bands observed in NH stretching region (3224 cm$^{-1}$ and 3268 cm$^{-1}$ in IR), are weaker than the doublets of doublets, and they must originate from vibrations of terminal [NH$_3$] groups of B3N3. Indeed, they fall in a spectral region typical for terminal [NH$_3$] groups of ammonia borane (3196 cm$^{-1}$, 3253 cm$^{-1}$ and 3311 cm$^{-1}$). Naturally, it is expected that signals originating from the terminal [NH$_3$] of B3N3 are weaker than those from more numerous [NH$_2$] groups present in both (NH$_4$)(B3N2) and B3N3.

The region of the IR absorption spectrum associated with deformations of the NH$_x$ moieties (Figure 7) is consistent with these conclusions. One can clearly distinguish signals at *ca.* 1530–1600 cm$^{-1}$, typical for [NH$_2$] and [NH$_3$] groups,[21] from the signals at *ca.* 1250–1500 cm$^{-1}$, characteristic for ammonium cations (*cf.* 1402 cm$^{-1}$ for NH$_4$Cl).[33] It is worth to notice, that the IR spectra in the 1350–1500 cm$^{-1}$ region show five bands (1374 cm$^{-1}$, 1392 cm$^{-1}$, 1415 cm$^{-1}$, 1427 cm$^{-1}$, 1479 cm$^{-1}$), and this agrees with the number of deformation modes expected for NH$_4^+$ cation in a low-symmetry environment.

The spectroscopic analysis clearly shows that both (NH$_4$)(B3N2) and B3N3 moieties constitute the main product, thus confirming the reactions according to Eq. 1 and Eq. 2.

### 2.4 Crystal structure of the side product

The chemical composition of the side product of metathesis was confirmed by single crystal x-ray diffraction measurements (*cf.* ESI). This compound contains protonated Verkade's base cations and tetraphenylborate anions, (VBH)[B(C$_6$H$_5$)$_4$], proving successful ion exchange in reaction according to Eq. 1. The compound crystallises in $P\bar{1}$ space group with the constituent ions of different polarities showing no significant interactions, as expected for large ions with small charge smeared over the entire ion.

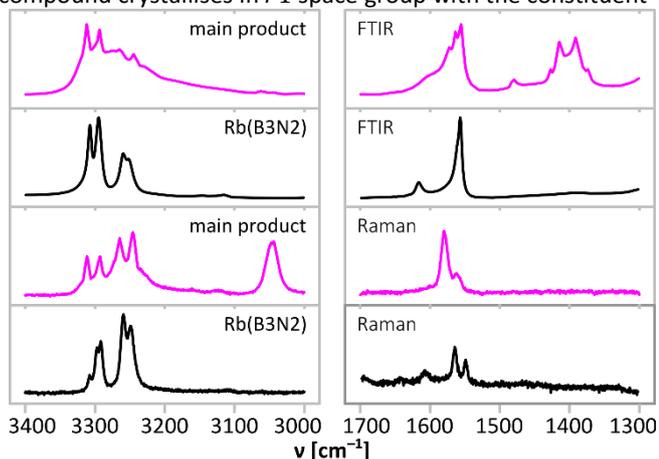

**Fig. 7.** Comparison of NH stretching (3000–3400 cm$^{-1}$) and NH bending (1300–1700 cm$^{-1}$) regions of FTIR absorption (top in each bracket) and Raman scattering (bottom in each bracket) spectra of the main product of the synthesis according to Eq. 1 (magenta) and alkali metal M(B3N2) salts. Full spectra presented in Fig. 6.

### 2.5 Crystal structure of the main product

As we could not obtain a single crystal of the main product, we were forced to use powder X-ray diffraction (PXRD), supported by DFT calculations and the results of spectroscopic analysis described above.

Indexing of the PXRD pattern leads to a $P2_1/c$ unit cell with the refined lattice parameters of: a = 13.391(10) Å, b = 13.195(8) Å, c = 17.822(12) Å, β = 125.86(4)° and V = 2552(3) Å$^3$. Assuming (NH$_4$)(B3N2) as a product, such unit cell volume would suggest Z = 16 (multiplicity of the general atomic position) and V/Z = 159.5 Å$^3$. However, this V/Z value is too small as the values for K and Rb analogues are larger (167.7 Å$^3$ and 174.3 Å$^3$ respectively)[21] while the size of NH$_4^+$ falls between these two alkali metal cations. Somewhat The smaller than expected V/Z volume suggests that the crystalline phase should contain also the partially dehydrogenated molecules of the product of condensation presented in Eq. 2.

To test such scenario, structural models were derived for (NH$_4$)(B3N2), B3N3 and the (NH$_4$)(B3N2)·3(B3N3) cocrystal with the components in the 1:3 molar ratio as indicated by NMR data. Initial positions of heavy atoms came from simulated annealing using the experimental diffraction data. The models were then fully optimized using periodic DFT calculations (Table 2 and SI). The theoretical unit cell volume calculated for (NH$_4$)(B3N2) is significantly larger than those for the models containing B3N3 moieties and the latter are only 4.0–4.5% larger than the experimental value. This degree of overestimation is rather typical for the GGA calculations.

Importantly, the closest H⋯H contacts in the optimized structure of (NH$_4$)(B3N2) remain unreasonably short (1.40 Å), and outside the distribution observed experimentally for the dihydrogen bonds (usually >1.80 Å), Fig. 9. This reconfirms that the main product is not a pure (NH$_4$)(B3N2). The minimum H⋯H contacts in the optimized crystal structures containing B3N3 are significantly longer (1.60–1.68 Å), and closer to typical values for very strong dihydrogen interactions. Therefore, we have used a theoretical model of the (NH$_4$)(B3N2)·3(B3N3) cocrystal to refine its crystal structure using the best experimental dataset, Figure 8 and Figure S9.3 (SI).

The obtained structural model of (NH$_4$)(B3N2)·3(B3N3) contains four formula units in the unit cell (Z = 4) with one asymmetric unit (Z' = 1).

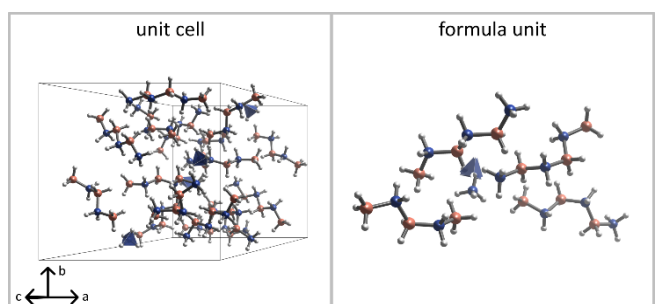

**Fig. 8.** Visualisation of the unit cell (left) and the asymmetric unit (right) of the crystal structure of the cocrystal comprising one unit of (NH$_4$)(B3N2) salt and three independent units of B3N3 molecules. Atom code: nitrogen – blue, boron – green, hydrogen – white.

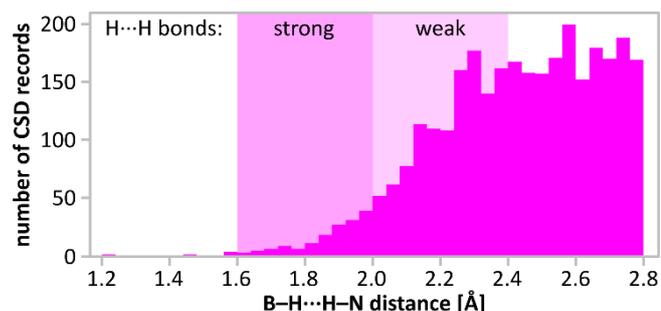

**Fig. 9.** Distribution of H··H distances in -B-H···H-N- moieties found in CSD database search (accessed by the end of 2020). The distances <2.8 Å were found for 865 crystal structures. Note the distances: 2.4 Å (double H van der Waals radius)[1] and 1.92 Å present in the structure.

The structure is stabilized by strong dihydrogen interactions. Two B3N3 chains adopt gauche geometry, and resemble the B3N2 anions in the heavier M(B3N2) salts, M = Rb, Cs.[21] The third B3N3 moiety and (B3N2)⁻ anion are more straight, closer to the geometry of anionic moieties in the light M(B3N2) salts, M = Li–K.[21,22] The B–N distances of 1.56(2)–1.58(2) Å remain within range observed in other compounds from this group. Further improvements of our preliminary experimental structure model, and in particular positions of hydrogen atoms, would require application of neutron diffraction methods, and is beyond the scope of this work.

**2.6 Thermal decomposition**

The theoretical gravimetric hydrogen content of the 1:3 cocrystals is very large, 16.4%. We have studied thermal decomposition of this new compound to assess its hydrogen storage properties. In Fig. 10 we present the results of a simultaneous thermogravimetric and calorimetric analysis of the main product together with gas evolution curves (hydrogen, ammonia, diborane, borazine) acquired in mass spectrometry experiment of the evolved gases.

The main product is thermally stable below 50°C. At higher temperatures, one can observe a multistep exothermic decomposition preceded by an endothermic process. During decomposition, a mixture of gases containing borazine, hydrogen, diborane and ammonia is being evolved. Close analysis of TGA/DSC/MS curves suggests that decomposition proceeds in at least 3 steps below 200°C, but each of them seems to have a similar profile of evolved gases. Interestingly, borazine is the main gaseous product of thermal decomposition, just as in the case of $NH_3BH_3$,[6] but dissimilarly to alkali metal M(B3N2) salts.[21,22] Facile evolution of borazine may be related to stoichiometry of two components of the main product. $(NH_4)(B3N2)$ and B3N3, both contain 3 boron atoms and 3 nitrogen atoms, just like borazine molecules. Dehydrogenation of B3N3 molecule proceeds with formation of a new B–N bond at [$BH_3$] and [$NH_3$] terminals according to Eq. 4. This reaction was also proposed as the final step of borazine evolution during thermal decomposition of ammonia borane.[34] Formation of pseudo-aromatic borazine is accompanied by dehydrogenation of the elusive head-to-tail cyclohexane-like intermediates, according to Equation 4.

**Table 2.** Summary of the DFT results. Minimum H…H distance is given for fully optimized unit cell and those which lattice vectors were fixed at experimental values. The model of $(NH_4)(B3N2)\cdot3(B3N3)$ refined to the experimental XRD data is added for comparison.

| Compound | V [Å³] | ΔV [%] | d(H…H)$_{min}$ cell opt. [Å] | d(H…H)$_{min}$ cell fix [Å] |
|---|---|---|---|---|
| $NH_4(B3N2)$ | 2832.0 | 11.0 | 1.40 | 1.42 |
| $(NH_4)(B3N2)\cdot3(B3N3)$ | 2666.1 | 4.5 | 1.60 | 1.62 |
| | *2552.2(30)** | – | *1.92** | – |
| B3N3 | 2654.7 | 4.0 | 1.68 | 1.65 |

*\* Experimental data with the lower constrain on the H…H separation.*

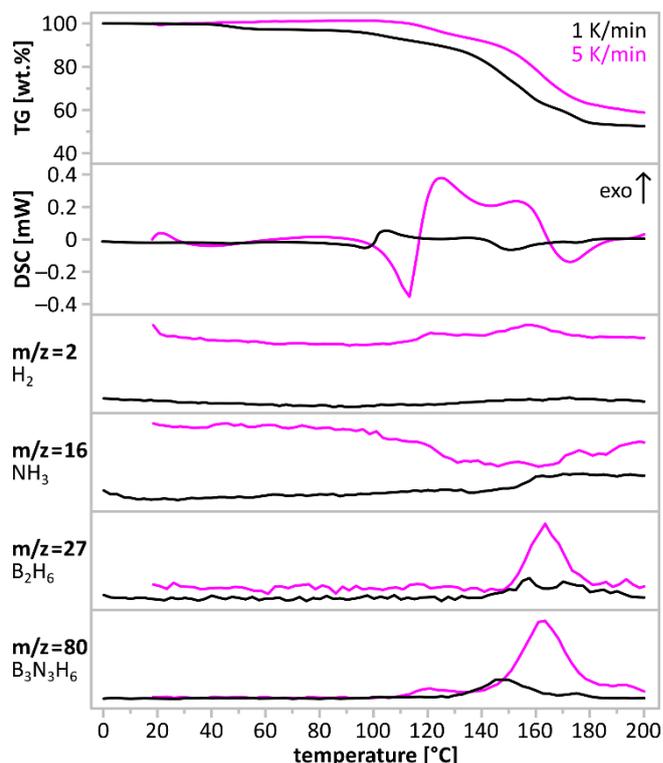

**Fig. 10.** Thermogravimetric (TG) and calorimetric (DCS) profiles of the main product set together with MS profiles of the evolved volatile products: hydrogen (m/z = 2), ammonia (m/z = 16), diborane (m/z = 27), borazine (m/z = 80). Scanning rate: 1 K/min (black), 5 K/min (magenta).

$NH_3BH_2NH_2BH_2NH_2BH_3 \rightarrow c\text{-}N_3B_3H_{12} + H_2 \rightarrow B_3N_3H_6 + 4H_2\uparrow$ (4)

We also observed the formation of $B_2H_6$ and $NH_3$, similarly as in decomposition of $Na(B3N2)$,[21,22] $K(B3N2)$,[21] $Rb(B3N2)$[21] and $Cs(B3N2)$,[21] which come from fragmentation of $(B3N2)^-$ anions.

As mentioned above, the thermal decomposition of the main product is preceded by an endothermic process. Analogies to ammonia borane[6] and amidoboranes[10] might suggest melting of the sample. However, direct visual observations ruled out this possibility. Endothermic event is related either to an intermolecular reorganization or a structural phase transition.

Depending on the heating rate, decomposition temperature reaches the fastest rate at *ca.* 145°C *ca.* 152°C for 1 K/min and 5 K/min scans, respectively. The observed mass loss upon thermal decomposition in the range up to 200°C, equals *ca.* 45% and surpasses those of alkali metal M(B3N2) salts[21] and parent ammonia borane.[6] Such large observed mass loss may be attributed to the evolution of borazine and other volatiles. The solid residue is amorphous, and consists of boron nitride and polymeric $B_XN_YH_Z$ phases as deduced from FTIR analysis (*cf.* ESI).

The observations discussed above lead to the following overall equations describing thermal decomposition of the two components of the main product:

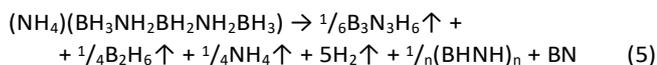

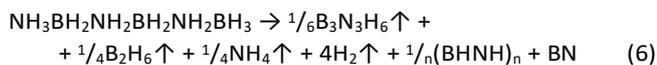

Theoretical mass loss of Eq. 5 and Eq.6 equal 43% and 42%, respectively, which reasonably agree with the observed experimental mass loss of *ca.* 45%.

## Conclusions

Synthesis of hydrogen-rich $(NH_4)(B3N2)$ salt was attempted in metathetic approach using precursors which contained weakly coordinating ions. The obtained product, however, corresponds to a mixture of ionic $(NH_4)(B3N2)$ and neutral B3N3 forming cocrystals in molar ratio of 1:3. Based on available $^{11}B$ NMR data, the main product was found to be very similar to the samples reported earlier by Ewing et al.[19] as B3N3.

The compound crystallises in $P2_1/c$ unit cell with the lattice parameters of: a = 13.401(11) Å, b = 13.196(8) Å, c = 17.828(12) Å, β = 128.83(4)°, V = 2556(3) Å$^3$ and Z = 16. The expected side product of a metathetic reaction, $(VBH)[B(C_6H_5)_4]$, crystallises in $P\bar{1}$ unit cell with the lattice parameters of a = 11.7376(3) Å, b = 19.5388(5) Å, c = 20.5479(4) Å, α = 61.751(2)°, β = 73.618(2)°, γ = 89.605(2)°, V = 3937.71 Å$^3$ and Z = 4.

Despite its high hydrogen content of 16.4%, the new compound cannot act as a self-standing solid-state hydrogen reservoir as it decomposes via a set of exothermic events while evolving mixture of volatile gaseous products such as borazine, diborane and ammonia, aside from hydrogen. However, it is possible that templating our product in porous matrixes could result in substantial improvement of purity of the evolved hydrogen, similarly as it was observed for ammonia borane.[35]

## Experimental

**Reagents:** All operations were performed under inert Ar atmosphere inside gloveboxes, MBRAUN Labmaster DP or Vigor SG1200 ($O_2$, $H_2O$ < 1.0 ppm). Commercially available reagents and solvents were used: $NH_3BH_3$ (98%, JSC Aviabor), $NH_4B(C_6H_5)_4$ (99%, Sigma-Aldrich (later denoted as SA), $C_4H_8O$ (99%, SA), $CH_2Cl_2$ (99%, SA). The synthesis of $(C_{18}H_{39}N_4PH)(BH_3NH_2BH_2NH_2BH_3)$ was performed according to the route described in our earlier paper.[21] For NMR measurements we used THF-d8 (99.5 atom% D, SA).

**Infrared absorption spectroscopy:** FTIR spectra were measured in the standard range of 400–4000 cm$^{-1}$ using Fourier Transform IR spectrometer Vertex 80v from Bruker. Samples were examined using KBr pallets prepared using anhydrous KBr (99%, SA) additionally dried in 150°C for 24h.

**Raman spectroscopy**: Raman scattering measurements were done using Raman microscopy setup from Jobin Yvon T64000 with Si CCD detector and Kr-Ar gas laser from Spectraphysics. We used green 514.5 nm excitation line. For the measurements small doses of samples were placed in 0.5 mm thick quartz capillaries sealed under inert gas atmosphere.

**Nuclear Magnetic Resonance:** $^1B$ NMR spectra with and without $^1H$ decoupling were obtained using Agilent 700 MHz spectrometer with Direct Drive 2 console and 5 mm room-temperature broadband probe. We used deuterated tetrahydrofuran (d$_8$-THF) as a solvent. The number of scans has been set to 256, the interscan delay to 1 s and the acquisition time to 200 ms. The spectra were acquired at 25 C. The exponential apodization has been used during processing (line broadening of 5 Hz).

**Thermogravimetrical Analysis:** Thermal decomposition was investigated using STA 410 thermal analyser from Netzsch, in the temperature range from −10°C to +200°C. STA 449 allows simultaneous thermogravimetric analysis, differential scanning calorimetry and evolved gas analysis by means of mass spectrometry. The samples were loaded into alumina crucibles inside a glovebox. Helium was used as a carrier gas. Evolved gases were analysed with a QMS 403C Aëolos MS from Pfeiffer–Vacuum. Transfer line was preheated to 100°C to avoid condensation of residues.

**Powder X-ray diffraction:** PXRD measurements were conducted on samples sealed in 0.5 mm thick quartz capillaries under inert atmosphere. Two diffractometers were used: Panalytical X'Pert Pro with linear PIXcel Medipix2 detector (parallel beam; the CoKα$_{12}$ radiation); and Bruker D8 Discover with 2D Vantec detector (parallel beam; the CuKα$_{12}$ radiation).

**Crystal structure solution of the main product:** Diffraction signals have been indexed using X-cell[36] and the initial structural model has been obtained using FOX software,[37] while the Rietveld refinement has been performed in Jana2006.[38] Pseudovoight functions with Berrar-Baldinozzi asymmetry have been used for modeling of diffraction profiles. The restraints were used during refinement for the N–H and B–H distances (at 0.900(10) Å and 1.100(10) Å, respectively, Fig. S9.2), and the angles related to hydrogen atoms (to 109.47° with tolerance of ca. 0.5°). The N–B distances were set to the value 1.57(1) Å. The atomic displacement parameters of B and N atoms were set equal, while those of H atoms were constrained according to the riding model. The bottom constraint of 1.91 Å for H…H distances was applied. Further details on the crystal structure may be obtained from CCDC/FIZ Karlsruhe on quoting the CSD deposition No. 2193624.

**Crystal structure solution of the side product:** Crystal of the compound were covered with perfluorinated oil (Krytox 1531). Data collection and reduction was performed with Agilent Supernova X-ray diffractometer with Kα-Cu radiation (microsource) with data reduction performed by CrysAlisPro software (v. 40.99).[39] Structure solution: SHELXT,[40] refinement against F$^2$ in Shelxl-2018, with ShelXle as GUI software.[41] The disorder of the –OC(CF$_3$)$_3$ groups was resolved using DSR.[42] Further details on the crystal structure may be obtained from CCDC/FIZ Karlsruhe on quoting the CSD deposition No. XXX.

**Density Functional Theory** (DFT) calculations were performed using CASTEP.[43] Generalized Gradient Approximation (GGA) was used with PBE functional and

Tkatchenko-Scheffler dispersive correction.[44] The cutoff value of 500 eV was applied to achieve good energy convergence. The density of the k-point grid was set below 0.1 Å$^{-1}$ and ultrasoft, generated on the fly pseudopotentials were used as they provide more accurate lattice parameters.

**Graphical presentation** of crystal structures has been performed with Vesta.[45]

## Conflicts of interest

There are no conflicts of interests to declare.

## Acknowledgements

This research was funded by Polish National Science Centre within the projects Preludium 13 (UMO/2017/25/N/ST5/01977) and Sonata Bis 8 (UMO/2018/30/E/ST5/00854). Research was carried out with the use of CePT infrastructure financed by the European Union – the European Regional Development Fund within the Operational Programme "Innovative economy" for 2007–2013 (POIG.02.02.00-14-024/08-00).

## Notes and references

# ELECTRONIC SUPLEMENTARY INFORMATION

"Towards hydrogen-rich ionic (NH$_4$)(BH$_3$NH$_2$BH$_2$NH$_2$BH$_3$) and related molecular NH$_3$BH$_2$NH$_2$BH$_2$NH$_2$BH$_3$"

R. Owarzany, T. Jaroń, K. Kazimierczuk, P. J. Malinowski, W. Grochala, K. J. Fijalkowski

**Contents:**

1. Records of reports on novel amidoborane and M(B3N2) salts and related compounds:

2. Synthesis of M(B3N2) salts of the products obtained:
   (NH$_4$)(B3N2)·3(B3N3)   Li(B3N2)   Na(B3N2)   K(B3N2)   Rb(B3N2)   Cs(B3N2)   NH$_3$BH$_3$

3. Table of $^{11}$B NMR @ THF-d$_8$ chemical shifts of M(B3N2) salts and ammonia borane:
   (NH$_4$)(B3N2)·3(B3N3)   Li(B3N2)   Na(B3N2)   K(B3N2)   Rb(B3N2)   Cs(B3N2)   NH$_3$BH$_3$   VBH(B3N2)

4. Table of bands appearing in the FTIR spectra of M(B3N2) salts and ammonia borane:
   (NH$_4$)(B3N2)·3(B3N3)   Li(B3N2)   Na(B3N2)   K(B3N2)   Rb(B3N2)   Cs(B3N2)   NH$_3$BH$_3$

5. Table of bands appearing in the Raman spectra of M(B3N2) salts and ammonia borane:
   (NH$_4$)(B3N2)·3(B3N3)   Li(B3N2)   Na(B3N2)   K(B3N2)   Rb(B3N2)   Cs(B3N2)   NH$_3$BH$_3$

6. Comparison FTIR and Raman spectra of M(B3N2) salts:
   (NH$_4$)(B3N2)·3(B3N3)   Li(B3N2)   Na(B3N2)   K(B3N2)   Rb(B3N2)   Cs(B3N2)

7. Thermal decomposition (TGA curves) of M(B3N2) salts:
   (NH$_4$)(B3N2)·3(B3N3)   Li(B3N2)   Na(B3N2)   K(B3N2)   Rb(B3N2)   Cs(B3N2)

8. FTIR spectra of the products of thermal decomposition of M(B3N2) salts:
   (NH$_4$)(B3N2)·3(B3N3)   Li(B3N2)   Na(B3N2)   K(B3N2)   Rb(B3N2)   Cs(B3N2)

9. Experimental crystal structure and Rietveld fit for (NH$_4$)(B3N2)·3(B3N3):
   (NH$_4$)(B3N2)·3(B3N3)

10. Table with the closest H···H distances in the crystal structure of (NH$_4$)(B3N2)·3(B3N3):
    (NH$_4$)(B3N2)·3(B3N3)

11. Experimental and modelled NMR spectra for various possible compositions of the main product:
    (NH$_4$)(B3N2)·3(B3N3)   (NH$_4$)(B3N2)   (B3N3)

12. Results of DFT optimisation of modelled crystal structures:
    (NH$_4$)(B3N2)·3(B3N3)   (NH$_4$)(B3N2)   (B3N3)

13. Crystal structure (VBH)[B(C$_6$H$_5$)$_4$]
    (VBH)[B(C$_6$H$_5$)$_4$]

# 1. Records of reporting synthesis of novel amidoborane and M(B3N2) salts and related compounds:

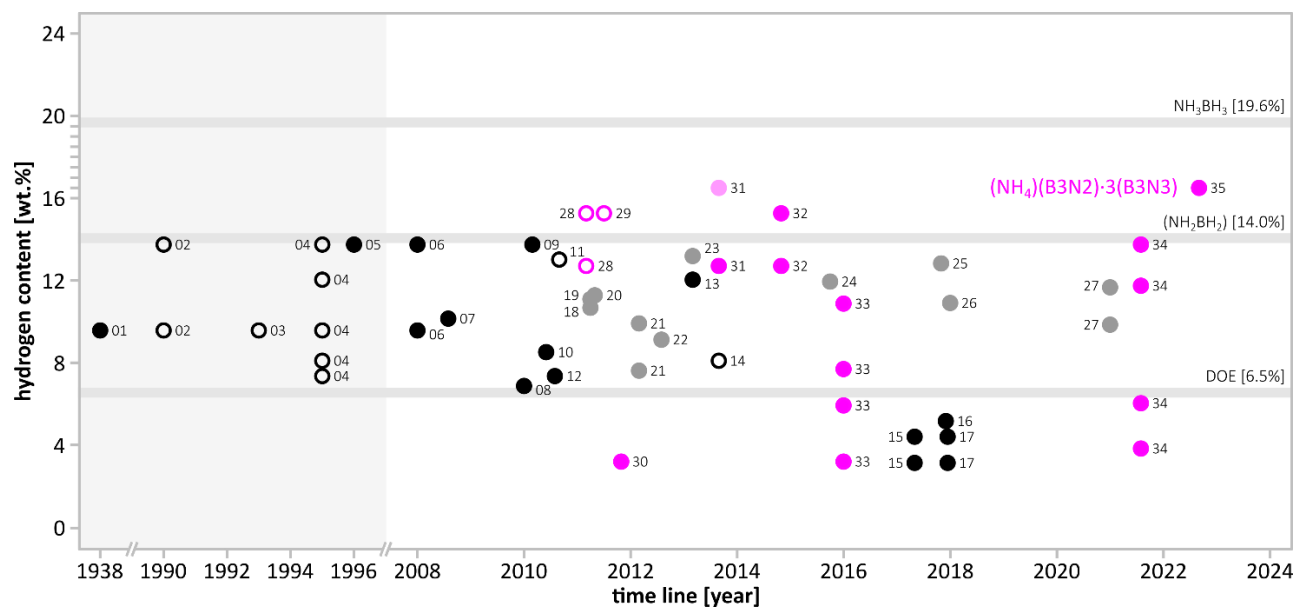

**Fig. S1.1.** Hydrogen content of monometallic amidoboranes (black), bimetallic amidoboranes (grey), M(B3N2) salts (magenta) and as a function of reporting date. Hydrogen content of $NH_3BH_3$ (19.6%), polymeric $(NH_2BH_2)$ (14.0%) and DOE ultimate target (6.5%) given as a reference. Reports and theses marked with hollow circles.

## 2. Synthesis of alkali metal M(B3N2) salts:

All operations were performed under inert Ar atmosphere inside gloveboxes, MBRAUN Labmaster DP or Vigor SG1200 ($O_2$, $H_2O$ < 1.0 ppm). Commercially available reagents and solvents were used: $NH_3BH_3$ (98%, JSC Aviabor), $NH_4B(C_6H_5)_4$ (99%, Sigma-Aldrich (later denoted as SA), $C_4H_8O$ (99%, SA), $CH_2Cl_2$ (99%, SA).

**Metathetic synthesis** was performed using $(C_{18}H_{39}N_4PH)(BH_3NH_2BH_2NH_2BH_3)$ and $NH_4[B(C_6H_5)_4]$ in anhydrous THF at room temperature under argon atmosphere:

$(C_{18}H_{39}N_4PH)(BH_3NH_2BH_2NH_2BH_3) + NH_4[B(C_6H_5)_4] \rightarrow (NH_4)(BH_3NH_2BH_2NH_2BH_3) + (C_{18}H_{39}N_4PH)[B(C_6H_5)_4]$

A follow-up process of dehydrogenation of $NH_4(BH_3NH_2BH_2NH_2BH_3)$ occurs leading to neutral linear molecule $NH_3BH_2NH_2BH_2NH_2BH_3$:

$(NH_4)(BH_3NH_2BH_2NH_2BH_3) \rightarrow NH_3BH_2NH_2BH_2NH_2BH_3 + H_2\uparrow$

Obtained mixture of products was well soluble in THF. Side product $(C_{18}H_{39}N_4PH)[B(C_6H_5)_4]$ was precipitated by washing with anhydrous DCM.

The main product crystallises in $P2_1/c$ unit cell with the lattice parameters of: a = 13.401(11) Å, b = 13.196(8) Å, c = 17.828(12) Å, β = 128.83(4)°, V = 2556(3) Å$^3$ and Z = 16. The crystalline product contains two compounds: $NH_4(BH_3NH_2BH_2NH_2BH_3)$ and $NH_3BH_2NH_2BH_2NH_2BH_3$ in molar ratio 1:3. In the manuscript, the product is denoted as "main product" or "$(NH_4)(B3N2) \cdot 3(B3N3)$".

The synthesis of $(C_{18}H_{39}N_4PH)(BH_3NH_2BH_2NH_2BH_3)$ was performed according to the route described in our earlier paper (R. Owarzany, *et al.*, *Inorg. Chem.* 55 (2016) 37/) in a direct reaction of Verkade's Base with 3 equivalents of ammonia borane in toluene at room temperature:

$C_{18}H_{39}N_4P + 3NH_3BH_3 \rightarrow (C_{18}H_{39}N_4PH)(BH_3NH_2BH_2NH_2BH_3) + H_2\uparrow + NH_3\uparrow$

**Different route of metathetic synthesis** between $Na(BH_3NH_2BH_2NH_2BH_3)$ and $NH_4Cl$ in glyme at room temperature for 24 hours was reported earlier (W. C. Ewing *et al. Inorg. Chem.* 52 (2013) 10690.), however, the authors were aiming $NH_3BH_2NH_2BH_2NH_2BH_3$ according to the following equation:

$Na(BH_3NH_2BH_2NH_2BH_3) + NH_4Cl \rightarrow NH_3BH_2NH_2BH_2NH_2BH_3 + H_2\uparrow + NaCl\downarrow$

Judging from the comparison of NMR data presented by Ewing *et al.* to our own data we strongly believe that this process leads to $(NH_4)(B3N2) \cdot 3(B3N3)$ according to the following reaction equations:

$Na(BH_3NH_2BH_2NH_2BH_3) + NH_4Cl \rightarrow (NH_4)(BH_3NH_2BH_2NH_2BH_3) + NaCl\downarrow$

$(NH_4)(BH_3NH_2BH_2NH_2BH_3) \rightarrow NH_3BH_2NH_2BH_2NH_2BH_3 + H_2\uparrow$

## 3. Table of [11]B NMR @ THF-d$_8$ chemical shifts of M(B3N2) salts and ammonia borane:

**Table S3.** Chemical shifts, positions of multiplets, excitation frequencies and J-coupling values observed in [11]B NMR spectra in deuterated THF solution (δ [ppm]) of (NH$_4$)(B3N2)·3(B3N3) at room temperature. Results for ammonia borane [AB], precursor [β-VBH(B3N2)] and alkali metal M(B3N2) salts: [Li(B3N2), Na(B3N2), K(B3N2), Rb(B3N2), Cs(B3N2)] at RT are shown for comparison.

|  | NH$_3$BH$_3$ | VBH(B3N2) | M(B3N2) salts | | | | | (NH$_4$)(B3N2)·3(B3N3) |
|---|---|---|---|---|---|---|---|---|
|  |  |  | Li(B3N2) | Na(B3N2) | K(B3N2) | Rb(B3N2) | Cs(B3N2) |  |
| BH$_2$ triplet | – | −6.590 | −6.743 | −7.155 | −7.499 | −7.410 | −7.792 | −9.95 \| −11.78 |
|  | – | −8.184 | −8.360 | −8.582 | −8.568 | −8.424 | −8.384 | −10.40 \| −12.26 |
|  | – | −9.716 | −9.966 | −10.227 | −9.591 | −9.491 | −9.042 | −10.85 \| −12.70 |
| position | – | −8.163 | −8.356 | −8.654 | −8.553 | −8.442 | −8.406 | −10.4 \| −12.3 |
| [1]J (B,H) | – | 100 Hz | 103 Hz | 99 Hz | 101 Hz | 100 Hz | 101 Hz | 101 Hz \| 102 Hz |
| freq. | 96.32 MHz | 64.16 MHz | 96.32 MHz | 96.32 MHz | 96.32 MHz | 96.32 MHz | 160.48 MHz | 224.62 MHz |

|  | NH$_3$BH$_3$ | VBH(B3N2) | M(B3N2) salts | | | | | (NH$_4$)(B3N2)·3(B3N3) |
|---|---|---|---|---|---|---|---|---|
|  |  |  | Li(B3N2) | Na(B3N2) | K(B3N2) | Rb(B3N2) | Cs(B3N2) |  |
| BH$_3$ quartet | −18.184 | −19.462 | −20.465 | −20.264 | −20.613 | −20.294 | −20.279 | −21.60 |
|  | −19.632 | −20.933 | −21.836 | −21.693 | −21.556 | −21.226 | −20.902 | −22.01 |
|  | −21.144 | −22.342 | −23.314 | −23.202 | −22.499 | −22.168 | −21.479 | −22.42 |
|  | −22.610 | −23.752 | −24.634 | −24.582 | −23.382 | −23.105 | −22.025 | −22.83 |
| position | −20.393 | −21.622 | −22.562 | −22.435 | −22.013 | −21.698 | −21.171 | −22.21 |
| [1]J (B,H) | 95 Hz | 91 Hz | 90 Hz | 91 Hz | 89 Hz | 90 Hz | 94 Hz | 91 HZ |
| freq. | 96.32 MHz | 64.16 MHz | 96.32 MHz | 96.32 MHz | 96.32 MHz | 96.32 MHz | 160.48 MHz | 224.62 MHz |

## 4. Table of bands appearing in the FTIR spectra of M(B3N2) salts and ammonia borane:

**Table S4.** Absorption bands detected in IR spectra (wavenumber [cm⁻¹]) of NH₄(B3N2)·3(B3N3) at room temperature. Results for ammonia borane [AB] and alkali metal M(B3N2) salts: [Li(B3N2), Na(B3N2), K(B3N2), Rb(B3N2), Cs(B3N2)] at RT. Absorption bands of ammonia borane at RT are shown for comparison. (ν = stretching, δ = deformation: bending and torsional modes).

| Band | NH₃BH₃ | M(B3N2) salts | | | | | (NH₄)(B3N2)·3(B3N3) |
|------|--------|---------|---------|---------|---------|---------|---------|
|      |        | Li(B3N2) | Na(B3N2) | K(B3N2) | Rb(B3N2) | Cs(B3N2) |  |
| ν(NH) | 3311 vs | 3310 s | 3302 vs | 3305 vs | 3308 m<br>3295 m | 3313 w<br>3287 m | 3306 vs<br>3288 vs<br><br>3268 sh |
|  | 3253 vs | 3273 m | 3256 m | 3261 m | 3261 w<br>3252 w | 3261 w<br>3235 m | 3259 m<br>3239 m<br><br>3223 sh |
|  | 3196 s |  |  |  |  |  |  |
| ν(BH) |  |  |  | 2420 sh |  |  | 2439 sh<br>2407 s |
|  |  | 2350 vs | 2364 s | 2352 m | 2390 sh<br>2346 s | 2389 sh<br>2329 m | 2357 s |
|  | 2347 vs | 2322 s | 2315 s |  |  |  | 2317 vs |
|  |  |  |  | 2304 s |  |  | 2302 sh |
|  | 2289 s | 2282 vs | 2286 vs | 2279 s | 2294 vs | 2291 vs |  |
|  |  | 2245 s |  | 2259 s | 2263 s<br>2248 sh | 2248 s | 2260 m |
|  |  |  |  | 2210 sh | 2204 sh | 2189 sh |  |
|  | 2118 m |  |  |  |  |  |  |
| δ(NH) | 1611 m |  |  |  | 1617 vw |  | 1604 sh<br>1572 w |
|  |  | 1571 vs | 1576 w | 1583 w | 1562 sh | 1579 vw<br>1565 w | 1564 m |
|  |  |  | 1556 m | 1568 m | 1557 m | 1557 vw | 1556 m |
|  |  |  |  |  |  |  | 1480 w<br>1428 w<br>1415 m<br>1392 m<br>1375 w |
| δ(BH) |  | 1283 m | 1248 m | 1244 m |  | 1259 m | 1262 s |
|  |  | 1226 s |  |  | 1233 s | 1231 s | 1244 s |
|  |  | 1201 s | 1199 vs | 1202 s | 1206 m | 1205 s | 1203 vs |
|  |  |  | 1175 m | 1194 sh | 1193 sh | 1188 s | 1179 s |
|  | 1163 vs | 1148 s |  | 1182 m | 1167 s | 1171 s | 1168 sh |
|  |  | 1135 m | 1129 vw | 1128 w | 1134 m | 1127 m | 1134 m |
|  |  |  |  |  |  |  | 1118 w |
|  | 1067 s |  | 1074 m | 1073 w | 1065 w | 1063 w | 1078 w |
|  |  | 1044 m | 1055 m | 1056 w | 1038 w | 1043 m | 1056 w |
|  |  | 1013 w | 999 w | 997 w | 1012 vw | 1001 w | 983 w |
| ν(BN) and other |  | 916 w | 893 vw | 893 vw | 901 vw | 902 w | 861 w |
|  |  |  |  | 875 w | 873 vw | 881 w |  |
|  |  | 874 vw | 870 w |  | 854 vw | 856 w |  |
|  |  | 799 vw | 785 vw | 781 vw | 811 vw | 791 vw | 804 vw |
|  |  |  |  | 727 vw | 727 vw | 723 vw | 749 w |
|  |  |  |  |  |  |  | 711 w |

## 5. Table of bands appearing in the RAMAN spectra of M(B3N2) salts and ammonia borane:

**Table S5.** Absorption bands detected in Raman spectra (wavenumber [cm⁻¹]) of NH₄(B3N2)·3(B3N3) at room temperature. Results for ammonia borane [AB] and alkali metal M(B3N2) salts: [Li(B3N2), Na(B3N2), K(B3N2), Rb(B3N2), Cs(B3N2)] at RT. Absorption bands of ammonia borane at RT are shown for comparison. (ν = stretching, δ = deformation: bending and torsional modes).

| Band | NH₃BH₃ | M(B3N2) salts | | | | | (NH₄)(B3N2)·3(B3N3) |
|---|---|---|---|---|---|---|---|
| | | Li(B3N2) | Na(B3N2) | K(B3N2) | Rb(B3N2) | Cs(B3N2) | |
| ν(NH) | | | | | 3304 sh | 3307 sh | |
| | 3314 m | 3314 m | 3302 s | 3306 s | 3293 w | 3303 w | 3307 m |
| | | | | | 3288 w | 3292 sh | 3288 m |
| | | | | | | 3279 w | |
| | 3253 vs | 3272 s | 3265 vs | 3263 vs | 3256 m | 3258 m | 3260 vs |
| | | | | | 3245 m | 3232 m | 3241 vs |
| | 3177 m | | | | | | |
| | | | | | | | 3041 s |
| ν(BH) | | 2418 vw | 2403 w | | | | 2475 sh |
| | 2378 vs | 2370 w | 2373 w | 2382 m | 2379 m | 2379 m | 2442 w |
| | | | | 2347 m | 2341 m | 2343 m | 2394 m |
| | | 2320 vs | 2322 m | | | | 2320 vs |
| | | | | 2301 s | 2291 vs | 2286 vs | 2261 s |
| | 2284 vs | 2282 m | 2275 s | 2274 s | | | |
| | | 2250 s | 2243 s | | 2243 vs | 2250 vs | |
| | | | 2214 sh | | | | |
| | | 2166 vw | | 2186 sh | 2192 sh | | |
| δ(NH) | | | | 1649 vw | 1608 vw | 1645 vw | |
| | 1598 m | | | 1585 vw | | | 1578 s |
| | 1583 m | 1567 m | 1539 w | 1569 vw | 1565 w | 1566 w | 1559 m |
| | | | 1519 vw | | 1550 w | 1534 w | |
| δ(BH) | | 1281 w | | | | | |
| | | 1259 vw | | | | | |
| | | 1226 w | | | | | |
| | | 1206 m | 1212 w | 1227 w | 1207 w | 1206 sh | 1201 w |
| | 1190 sh | | | 1193 w | 1188 m | 1183 m | 1184 w |
| | 1168 m | 1166 w | 1162 w | 1168 w | 1155 w | 1163 m | 1173 w |
| | | | 1132 w | | 1122 vw | | 1155 w |
| | 1069 vw | | 1047 vw | | 1056 vw | 1036 vw | 1102 w |
| | | | | | | | 1027 m |
| | | 1010 w | 1019 vw | | 1004 vw | 993 w | 998 vs |
| ν(BN) and other | | | | | | 915 w | |
| | | 895 vw | | 892 vw | 895 w | 899 vw | |
| | | 873 w | | 871 w | 862 w | 876 vw | |
| | | | 856 vw | | 847 w | 851 w | 857 w |
| | 800 w | 806 w | 835 w | | | | 796 w |
| | 785 m | | 749 w | 778 w | 779 w | 783 vw | |
| | 729 w | | | | 721 w | 724 w | |
| | | | | | | 715 vw | |
| | | | 614 vw | | 653 w | 643 vw | 667 m |
| | | | | | 639 vw | | 618 w |
| | | | | | | | 604 s |

## 6. Comparison of FTIR and Raman spectra of alkali metal M(B3N2) salts:

FTIR and Raman spectra of NH$_4$(B3N2)/3(B3N3) and alkali metal M(B3N2) salts: [Li(B3N2), Na(B3N2), K(B3N2), Rb(B3N2), Cs(B3N2)]. NH and BH stretching and NH bending regions highlighted and magnified in separate figures.

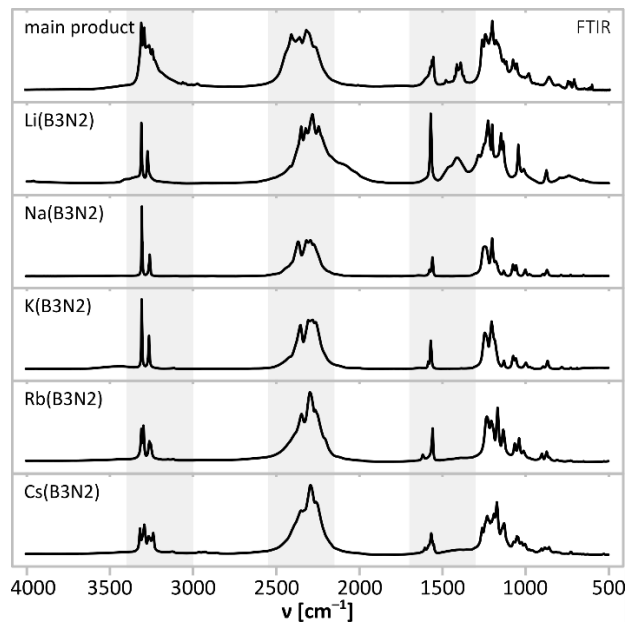

**Fig. S6.1.** Comparison of FTIR spectra of M(B3N2) salts.

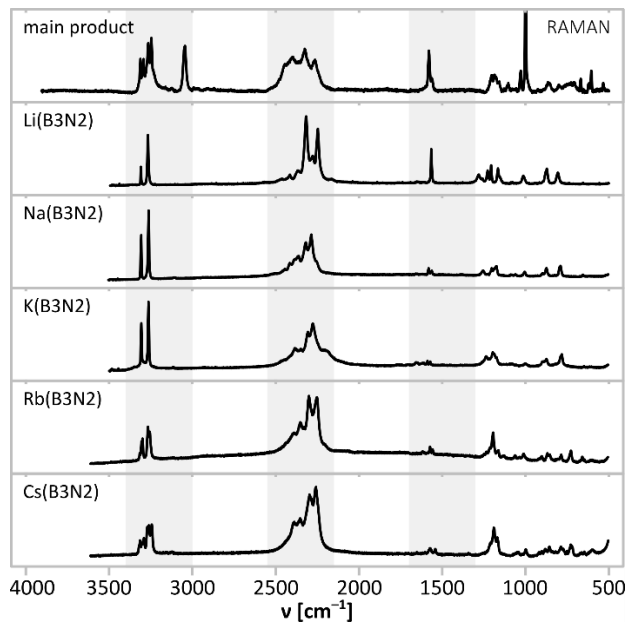

**Fig. S6.2.** Comparison of Raman spectra of M(B3N2) salts.

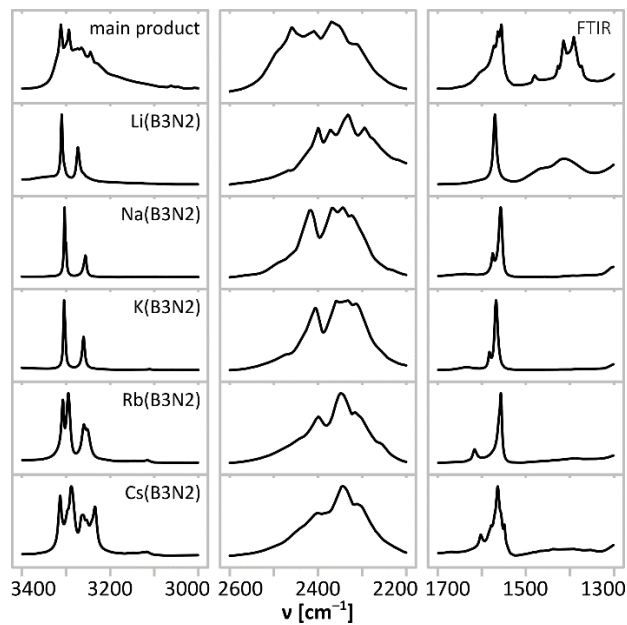

**Fig. S6.3.** Comparison of NH and BH stretching and NH bending regions of FTIR spectra of M(B3N2) salts.

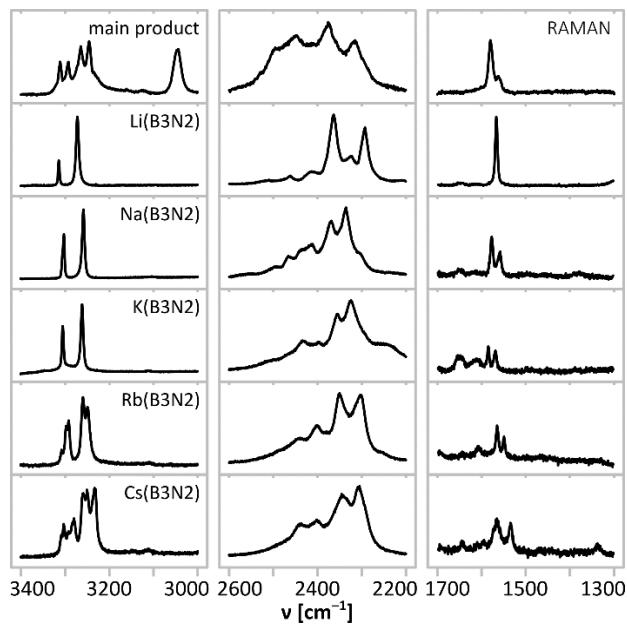

**Fig. S6.4.** Comparison of NH and BH stretching and NH bending regions of Raman spectra of M(B3N2) salts.

## 7. Thermal decomposition (TGA curves) of M(B3N2) salts:

The thermal decomposition of NH₄(B3N2)/3(B3N3) and alkali metal M(B3N2) salts occurs at the temperature range of 120–180°C.

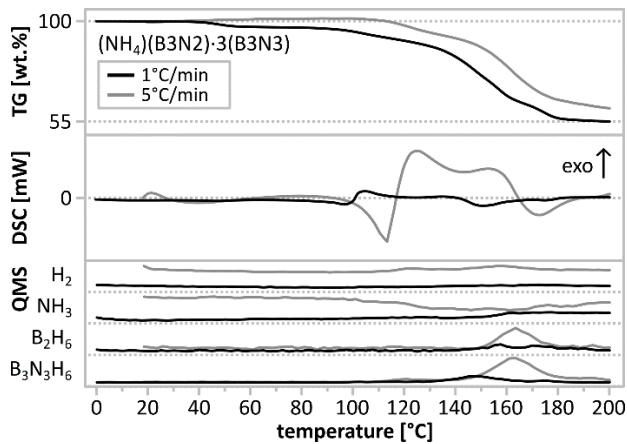

**Fig.S7.1.** TGA/DSC experiments of (NH₄)(B3N2)·3(B3N3) with scanning rates (1 K/min -black, 5 K/min -grey).

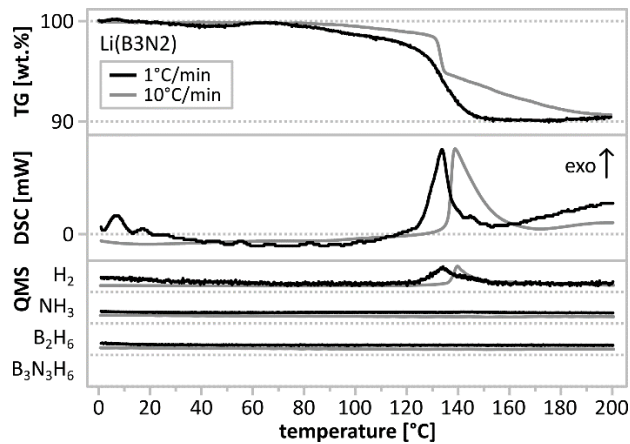

**Fig.S7.2.** TGA/DSC experiments of Li(B3N2) sample with different scanning rates (1 K/min -black, 10 K/min -grey).

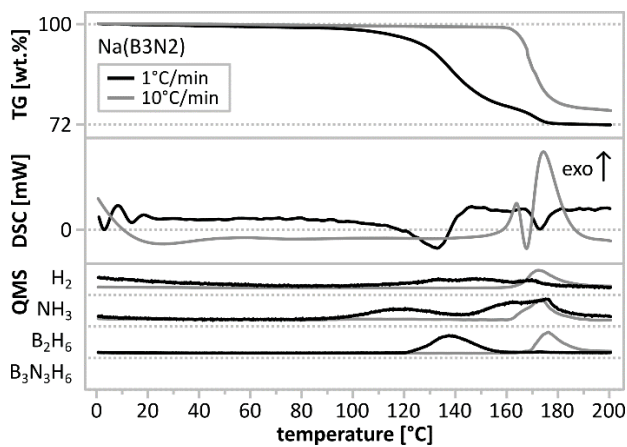

**Fig.S7.3.** TGA/DSC experiments of Na(B3N2) sample with different scanning rates (1 K/min -black, 10 K/min -grey).

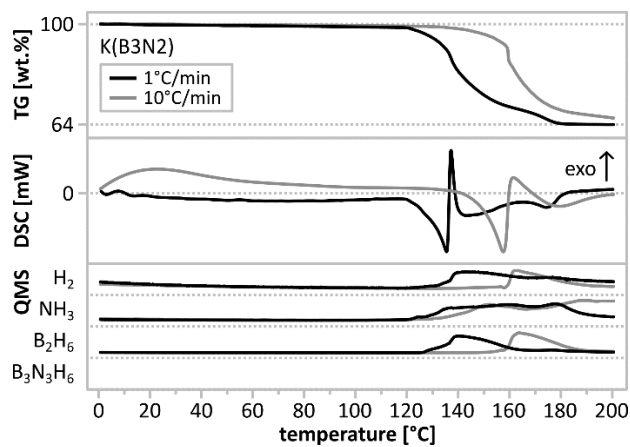

**Fig.S7.4.** TGA/DSC experiments of K(B3N2) sample with different scanning rates (1 K/min -black, 10 K/min -grey).

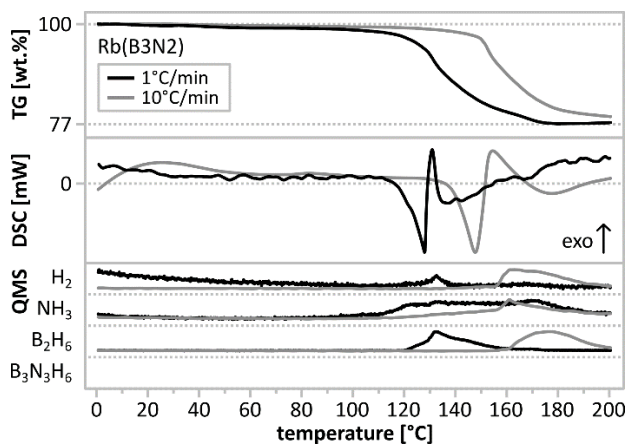

**Fig.S7.5.** TGA/DSC experiments of Rb(B3N2) sample with different scanning rates (1 K/min -black, 10 K/min -grey).

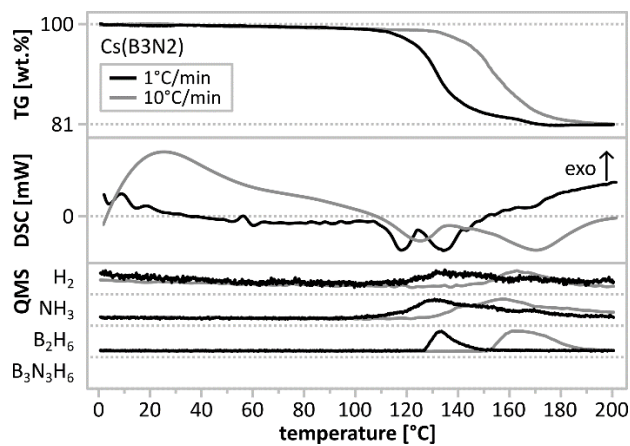

**Fig.S7.6.** TGA/DSC experiments of Cs(B3N2) sample with different scanning rates (1 K/min -black, 10 K/min -grey).

## 8. FTIR spectra of the products of thermal decomposition of M(B3N2) salts:

The thermal decomposition of (NH₄)(B3N2) leads to formation of boron nitride while decomposition of alkali metal M(B3N2) salts leads to formation of respective borohydrides.

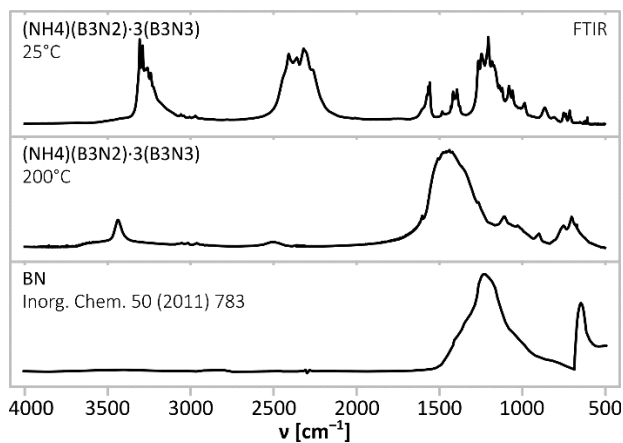

**Fig. S8.1.** FTIR spectra of the product of thermal decomposition of (NH₄)(B3N2)·3(B3N3) sample.

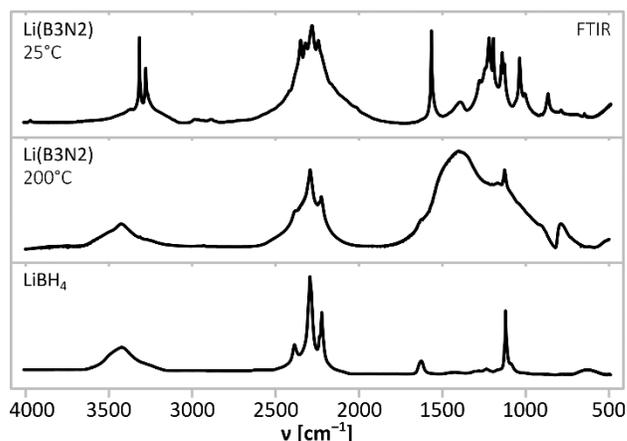

**Fig. S8.2.** FTIR spectra of the product of thermal decomposition of Li(B3N2) sample.

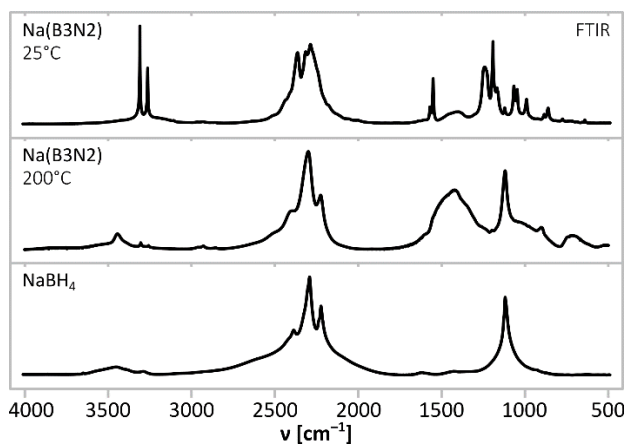

**Fig. S8.3.** FTIR spectra of the product of thermal decomposition of Na(B3N2) sample.

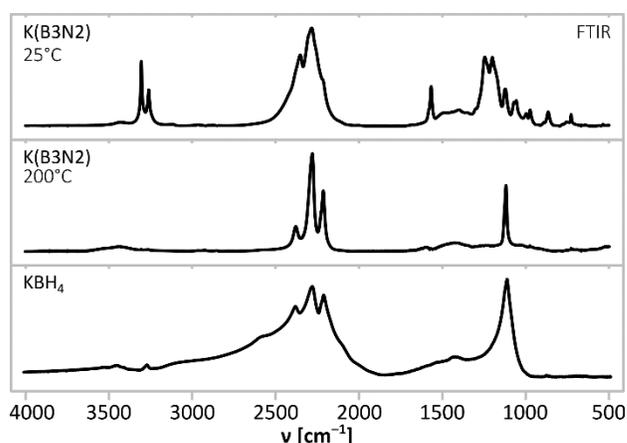

**Fig. S8.4.** FTIR spectra of the product of thermal decomposition of K(B3N2) sample.

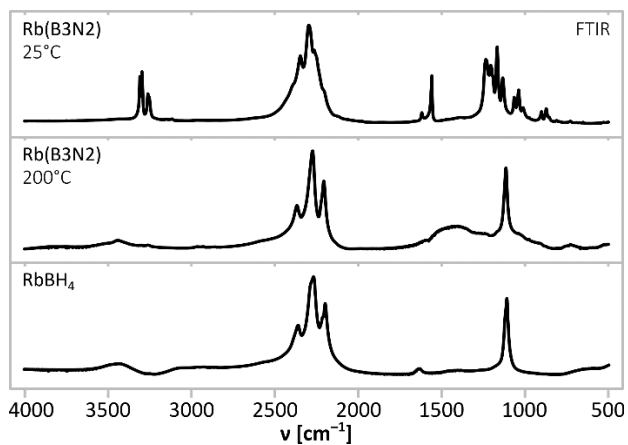

**Fig. S8.5.** FTIR spectra of the product of thermal decomposition of Rb(B3N2) sample.

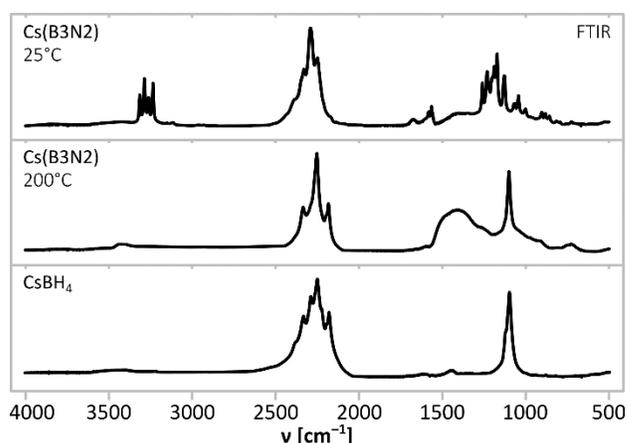

**Fig. S8.6.** FTIR spectra of the product of thermal decomposition of Cs(B3N2) sample.

## 9. Experimental crystal structure and Rietveld fit for (NH$_4$)(B3N2)·3(B3N3):

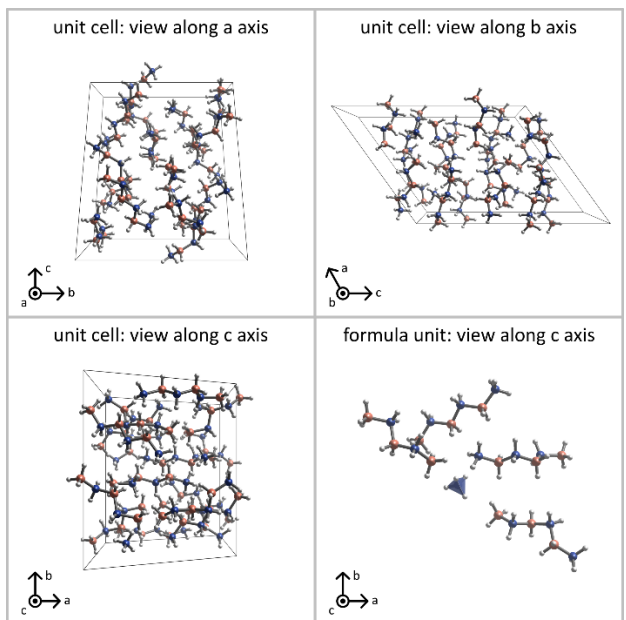

**Fig. S9.1.** Visualisation of the unit cell and formula unit of the main product: (NH$_4$)(B3N2)·3(B3N3)

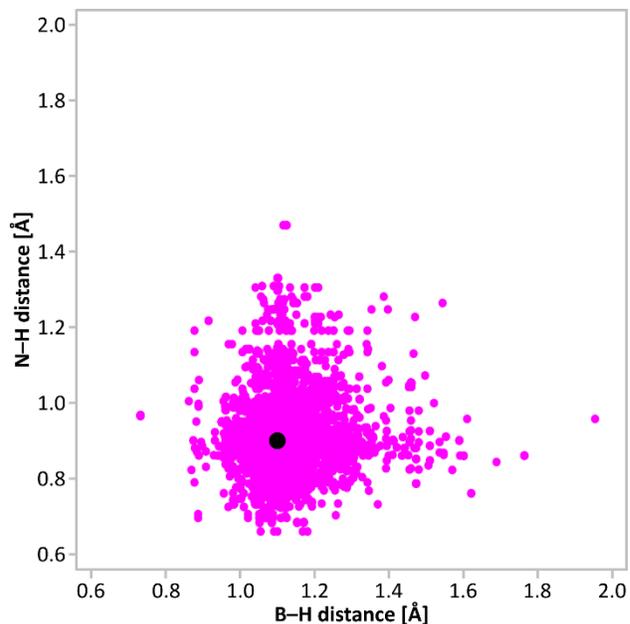

**Fig. S9.2.** Distribution of N-H and B-H distances in systems comprising both [NH$_x$] and [BH$_x$] groups found in structures in CSD database. Value of N-H and B-H distances in (NH$_4$)(B3N2)·3(B3N3) marked with a dot.

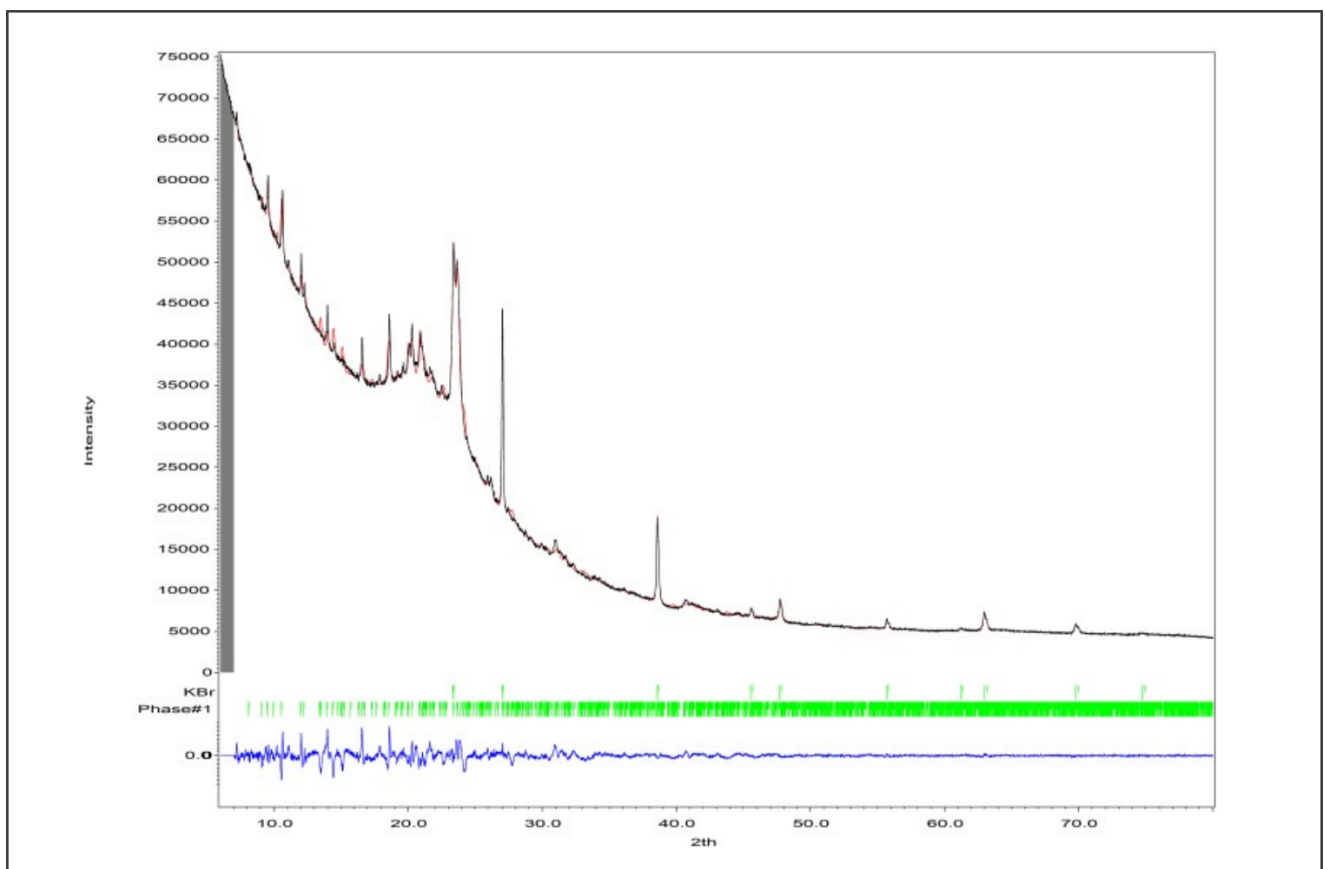

**Fig. S9.3.** Rietveld analysis of (NH$_4$)(B3N2)/3(B3N3) powder pattern. CoK$_{\alpha1,2}$, λ = 1.78901 Å.

## 10. Table with the closest H⋯H distances in the crystal structure of $(NH_4)(B3N2)·3(B3N3)$:

**Table S10.** List of the closest H⋯H distances in experimental crystal structure of $(NH_4)(B3N2)·3(B3N3)$. Listed only strong dihydrogen bonds, < 2 Å.

| H atom1 | H atom2 | Length [Å] | Length-VdW [Å] | Neighboring groups |
|---|---|---|---|---|
| H8  | H39 | 1.927 | -0.473 | B–H⋯H–N |
| H12 | H29 | 1.928 | -0.472 | B–H⋯H–N |
| H6  | H52 | 1.931 | -0.469 | B–H⋯H–N |
| H12 | H53 | 1.935 | -0.465 | B–H⋯H–N |
| H15 | H34 | 1.948 | -0.452 | B–H⋯H–B |
| H54 | H42 | 1.949 | -0.451 | B–H⋯H–N |
| H1  | H51 | 1.951 | -0.449 | B–H⋯H–N |
| H50 | H64 | 1.971 | -0.429 | N–H⋯H–N |
| H19 | H31 | 1.984 | -0.416 | B–H⋯H–B |
| H52 | H64 | 1.987 | -0.413 | N–H⋯H–N |
| H17 | H47 | 1.993 | -0.407 | B–H⋯H–N |

## 11. Experimental and modelled NMR spectra for various possible compositions of the main product:

NMR spectra were simulated for various discussed possible compositions of the main product to ease visual examination of the experimental spectra obtained by us and reported earlier by Ewing *et al.*, *Inorganic Chemistry* 52 (2013) 10690.

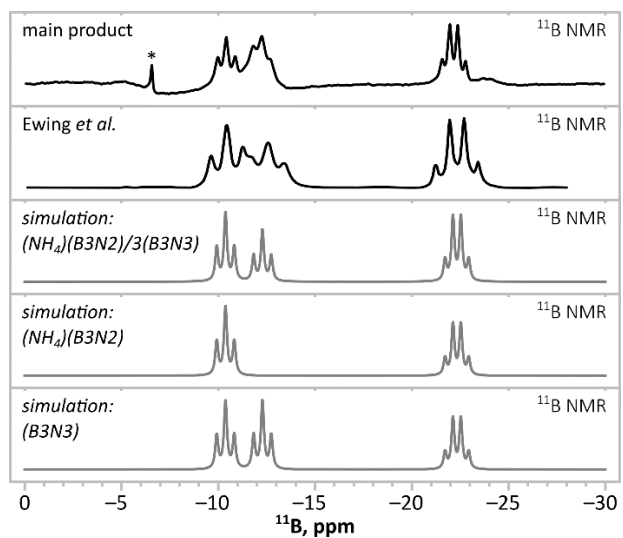
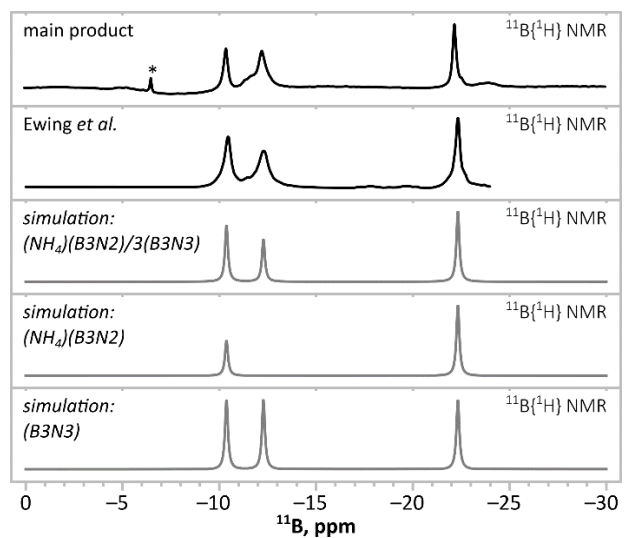

**Fig. S11.1.** Comparison of the experimental $^{11}$B NMR spectra obtained here and reported by Ewing *et al.* (*Inorganic Chemistry* 52 (2013) 10690) with spectra simulated for various possible compositions of the main product: (NH$_4$)(B3N2)·3(B3N3), (NH$_4$)(B3N2) and (B3N3).

**Fig. S11.2.** Comparison of the experimental $^{11}$B {$^1$H} NMR spectra obtained here and reported by Ewing *et al.* (*Inorganic Chemistry* 52 (2013) 10690) with spectra simulated for various possible compositions of the main product: (NH$_4$)(B3N2)·3(B3N3), (NH$_4$)(B3N2) and (B3N3).

## 12. Results of DFT optimisation of modelled crystal structures:

NH$_4$(BH$_3$NH$_2$BH$_2$NH$_2$BH$_3$)·3(NH$_3$BH$_2$NH$_2$BH$_2$NH$_2$BH$_3$), unit cell optimised
data_31_500eV\fine+cell
_audit_creation_date      2022-03-22
_audit_creation_method    'Materials Studio'
_symmetry_space_group_name_H-M   'P21/C'
_symmetry_Int_Tables_number    14
_symmetry_cell_setting    monoclinic
loop_
_symmetry_equiv_pos_as_xyz
 x,y,z
 -x,y+1/2,-z+1/2
 -x,-y,-z
 x,-y+1/2,z+1/2
_cell_length_a        13.2114
_cell_length_b        13.9512
_cell_length_c        19.0116
_cell_angle_alpha      90.0000
_cell_angle_beta      130.4617
_cell_angle_gamma      90.0000
loop_
_atom_site_label
_atom_site_type_symbol
_atom_site_fract_x
_atom_site_fract_y
_atom_site_fract_z
_atom_site_U_iso_or_equiv
_atom_site_adp_type
_atom_site_occupancy
B1   B   0.70821  0.11494  0.70797  0.05000  Uiso  1.00
N1   N   0.76629  0.14302  0.80732  0.05000  Uiso  1.00
B2   B   0.77307  0.25603  0.82169  0.05000  Uiso  1.00
N2   N   0.88332  0.27855  0.92649  0.05000  Uiso  1.00
B3   B   0.90995  0.38672  0.95751  0.05000  Uiso  1.00
H1   H   0.60523  0.15758  0.65051  0.05000  Uiso  1.00
H2   H   0.79337  0.12973  0.70200  0.05000  Uiso  1.00
H4   H   0.71177  0.11450  0.82381  0.05000  Uiso  1.00
H5   H   0.86071  0.11477  0.85440  0.05000  Uiso  1.00
H6   H   0.80153  0.29445  0.77863  0.05000  Uiso  1.00
H7   H   0.66525  0.28348  0.79429  0.05000  Uiso  1.00
H8   H   0.97286  0.25206  0.94872  0.05000  Uiso  1.00
H9   H   0.86665  0.24060  0.96452  0.05000  Uiso  1.00
H10  H   0.97588  0.38896  1.04154  0.05000  Uiso  1.00
H11  H   0.80194  0.42592  0.91975  0.05000  Uiso  1.00
H12  H   0.97000  0.42553  0.93710  0.05000  Uiso  1.00
B4   B   0.41303  0.33205  0.60553  0.05000  Uiso  1.00
N3   N   0.27925  0.35690  0.50627  0.05000  Uiso  1.00
B5   B   0.15136  0.33856  0.49619  0.05000  Uiso  1.00
N4   N   0.15353  0.41316  0.55956  0.05000  Uiso  1.00
B6   B   0.06225  0.39383  0.58541  0.05000  Uiso  1.00
H13  H   0.42249  0.24605  0.61536  0.05000  Uiso  1.00
H15  H   0.42037  0.37241  0.66568  0.05000  Uiso  1.00
H16  H   0.27593  0.42579  0.48531  0.05000  Uiso  1.00
H17  H   0.27253  0.31420  0.45908  0.05000  Uiso  1.00
H18  H   0.15917  0.25731  0.52314  0.05000  Uiso  1.00
H19  H   0.05145  0.35051  0.41580  0.05000  Uiso  1.00
H20  H   0.13000  0.48023  0.52964  0.05000  Uiso  1.00
H21  H   0.25125  0.41791  0.61940  0.05000  Uiso  1.00

```
H22  H  -0.05515  0.39939  0.51583  0.05000  Uiso  1.00
H23  H   0.08380  0.45473  0.64012  0.05000  Uiso  1.00
H24  H   0.08547  0.31319  0.61846  0.05000  Uiso  1.00
B7   B   0.35945  0.56131  0.01225  0.05000  Uiso  1.00
N5   N   0.39871  0.66288 -0.00050  0.05000  Uiso  1.00
B8   B   0.35089  0.76107  0.01297  0.05000  Uiso  1.00
N6   N   0.43801  0.78284  0.11929  0.05000  Uiso  1.00
B9   B   0.43903  0.89230  0.14365  0.05000  Uiso  1.00
H25  H   0.40686  0.49856 -0.00341  0.05000  Uiso  1.00
H26  H   0.39674  0.55325  0.08984  0.05000  Uiso  1.00
H28  H   0.50119  0.66593  0.04116  0.05000  Uiso  1.00
H29  H   0.36892  0.66512 -0.06589  0.05000  Uiso  1.00
H30  H   0.36881  0.82505 -0.02191  0.05000  Uiso  1.00
H31  H   0.23334  0.75618 -0.02405  0.05000  Uiso  1.00
H32  H   0.53513  0.76316  0.15269  0.05000  Uiso  1.00
H33  H   0.40908  0.74092  0.14813  0.05000  Uiso  1.00
H34  H   0.50214  0.93738  0.12797  0.05000  Uiso  1.00
H35  H   0.32535  0.92289  0.09508  0.05000  Uiso  1.00
H36  H   0.49200  0.89920  0.22589  0.05000  Uiso  1.00
B10  B   1.12914  0.99464  0.79230  0.05000  Uiso  1.00
N7   N   1.08247  0.88703  0.78744  0.05000  Uiso  1.00
B11  B   0.92947  0.86664  0.70636  0.05000  Uiso  1.00
N8   N   0.88287  0.77045  0.72226  0.05000  Uiso  1.00
B12  B   0.72463  0.75969  0.65604  0.05000  Uiso  1.00
H37  H   1.08482  1.04747  0.81782  0.05000  Uiso  1.00
H38  H   1.25109  0.99819  0.84578  0.05000  Uiso  1.00
H39  H   1.08213  1.01873  0.71408  0.05000  Uiso  1.00
H40  H   1.10997  0.87212  0.85076  0.05000  Uiso  1.00
H41  H   1.13739  0.84084  0.78158  0.05000  Uiso  1.00
H42  H   0.90914  0.85943  0.63407  0.05000  Uiso  1.00
H43  H   0.86774  0.93227  0.70464  0.05000  Uiso  1.00
H44  H   0.92197  0.71094  0.71553  0.05000  Uiso  1.00
H45  H   0.91808  0.76911  0.78909  0.05000  Uiso  1.00
H46  H   0.68750  0.80414  0.69155  0.05000  Uiso  1.00
H47  H   0.67311  0.79488  0.58006  0.05000  Uiso  1.00
H48  H   0.69326  0.67556  0.64739  0.05000  Uiso  1.00
N9   N   0.53664  0.37064  0.61627  0.05000  Uiso  1.00
H50  H   0.54045  0.44470  0.61701  0.05000  Uiso  1.00
H51  H   0.54125  0.34734  0.56683  0.05000  Uiso  1.00
H52  H   0.62319  0.34723  0.67920  0.05000  Uiso  1.00
N10  N   0.20232  0.54959 -0.06080  0.05000  Uiso  1.00
H53  H   0.15399  0.57777 -0.03889  0.05000  Uiso  1.00
H54  H   0.17501  0.47816 -0.07588  0.05000  Uiso  1.00
H55  H   0.15962  0.57929 -0.12419  0.05000  Uiso  1.00
N11  N   0.67472  0.00314  0.69163  0.05000  Uiso  1.00
H57  H   0.65387 -0.01839  0.63157  0.05000  Uiso  1.00
H58  H   0.75445 -0.03858  0.74368  0.05000  Uiso  1.00
H59  H   0.59194 -0.01451  0.68348  0.05000  Uiso  1.00
N12  N   0.23793  0.07365  0.68214  0.05000  Uiso  1.00
H61  H   0.21085  0.01630  0.63715  0.05000  Uiso  1.00
H62  H   0.19238  0.13480  0.64331  0.05000  Uiso  1.00
H63  H   0.34093  0.08208  0.72959  0.05000  Uiso  1.00
H64  H   0.20121  0.05757  0.71582  0.05000  Uiso  1.00
```

NH$_4$(BH$_3$NH$_2$BH$_2$NH$_2$BH$_3$), unit cell optimised
data_30_from00_noH-Hadd_500eV\fine+cell
_audit_creation_date       2022-03-22
_audit_creation_method        'Materials Studio'
_symmetry_space_group_name_H-M   'P21/C'
_symmetry_Int_Tables_number    14
_symmetry_cell_setting      monoclinic
loop_
_symmetry_equiv_pos_as_xyz
 x,y,z
 -x,y+1/2,-z+1/2
 -x,-y,-z
 x,-y+1/2,z+1/2
_cell_length_a         15.1296
_cell_length_b         14.0974
_cell_length_c         18.5157
_cell_angle_alpha        90.0000
_cell_angle_beta        134.1837
_cell_angle_gamma         90.0000
loop_
_atom_site_label
_atom_site_type_symbol
_atom_site_fract_x
_atom_site_fract_y
_atom_site_fract_z
_atom_site_U_iso_or_equiv
_atom_site_adp_type
_atom_site_occupancy
B1   B   0.66514  0.10694   0.69465  0.05000 Uiso  1.00
N1   N   0.71546  0.20346   0.75835  0.05000 Uiso  1.00
B2   B   0.85400  0.20822   0.86832  0.05000 Uiso  1.00
N2   B   0.91110  0.30099   0.95489  0.05000 Uiso  1.00
B3   N   1.01642  0.24767   1.06367  0.05000 Uiso  1.00
B4   B   0.58384  0.42106   0.65635  0.05000 Uiso  1.00
N3   N   0.48530  0.38527   0.65775  0.05000 Uiso  1.00
B5   B   0.43120  0.46721   0.67502  0.05000 Uiso  1.00
N4   N   0.30905  0.43511   0.64480  0.05000 Uiso  1.00
B6   B   0.19360  0.41548   0.52834  0.05000 Uiso  1.00
B7   B   0.23139  0.77848  -0.05771  0.05000 Uiso  1.00
N5   N   0.29779  0.79874   0.05453  0.05000 Uiso  1.00
B8   B   0.43371  0.76148   0.14552  0.05000 Uiso  1.00
N6   N   0.53044  0.80777   0.14794  0.05000 Uiso  1.00
B9   B   0.67184  0.78858   0.24664  0.05000 Uiso  1.00
B10  B   1.04876  0.97755   0.69603  0.05000 Uiso  1.00
N7   N   0.98926  0.87693   0.67794  0.05000 Uiso  1.00
B11  B   0.84730  0.88149   0.61679  0.05000 Uiso  1.00
N8   N   0.79452  0.78276   0.61159  0.05000 Uiso  1.00
B12  B   0.64739  0.78172   0.53724  0.05000 Uiso  1.00
N9   N   0.30421  1.02782   0.92396  0.05000 Uiso  1.00
N10  N   0.89991  0.38928   0.23582  0.05000 Uiso  1.00
N11  N   0.40411 -0.02701   0.53429  0.05000 Uiso  1.00
N12  N   0.16159  0.03439   0.59474  0.05000 Uiso  1.00
H25  H   0.57587  0.08165   0.67916  0.00000 Uiso  1.00
H26  H   0.74209  0.04326   0.74155  0.00000 Uiso  1.00
H27  H   0.63368  0.12074   0.61368  0.00000 Uiso  1.00
H28  H   0.65741  0.21604   0.76811  0.00000 Uiso  1.00
H29  H   0.69738  0.25979   0.71420  0.00000 Uiso  1.00
H30  H   0.92670  0.20886   0.85760  0.00000 Uiso  1.00
H31  H   0.87080  0.13194   0.90858  0.00000 Uiso  1.00

| | | | | | | | |
|---|---|---|---|---|---|---|---|
| H32 | H | 0.84498 | 0.34518 | 0.96014 | 0.00000 | Uiso | 1.00 |
| H33 | H | 0.96760 | 0.36208 | 0.95162 | 0.00000 | Uiso | 1.00 |
| H34 | H | 1.08626 | 0.28808 | 1.12530 | 0.00000 | Uiso | 1.00 |
| H35 | H | 1.05800 | 0.19576 | 1.05622 | 0.00000 | Uiso | 1.00 |
| H36 | H | 0.97446 | 0.21306 | 1.08218 | 0.00000 | Uiso | 1.00 |
| H37 | H | 0.54476 | 0.49423 | 0.60574 | 0.00000 | Uiso | 1.00 |
| H38 | H | 0.60038 | 0.36151 | 0.61857 | 0.00000 | Uiso | 1.00 |
| H39 | H | 0.67976 | 0.43811 | 0.74197 | 0.00000 | Uiso | 1.00 |
| H40 | H | 0.52471 | 0.33553 | 0.71369 | 0.00000 | Uiso | 1.00 |
| H41 | H | 0.41727 | 0.34701 | 0.59332 | 0.00000 | Uiso | 1.00 |
| H42 | H | 0.40227 | 0.53202 | 0.61842 | 0.00000 | Uiso | 1.00 |
| H43 | H | 0.51021 | 0.49321 | 0.76251 | 0.00000 | Uiso | 1.00 |
| H44 | H | 0.28544 | 0.48737 | 0.66769 | 0.00000 | Uiso | 1.00 |
| H45 | H | 0.32357 | 0.37640 | 0.68540 | 0.00000 | Uiso | 1.00 |
| H46 | H | 0.10247 | 0.39780 | 0.51231 | 0.00000 | Uiso | 1.00 |
| H47 | H | 0.21701 | 0.34798 | 0.50190 | 0.00000 | Uiso | 1.00 |
| H48 | H | 0.17298 | 0.48714 | 0.48003 | 0.00000 | Uiso | 1.00 |
| H49 | H | 0.23580 | 0.69217 | -0.06386 | 0.00000 | Uiso | 1.00 |
| H50 | H | 0.28267 | 0.81830 | -0.07846 | 0.00000 | Uiso | 1.00 |
| H51 | H | 0.12202 | 0.80402 | -0.11675 | 0.00000 | Uiso | 1.00 |
| H52 | H | 0.24488 | 0.76892 | 0.06495 | 0.00000 | Uiso | 1.00 |
| H53 | H | 0.29626 | 0.87047 | 0.06438 | 0.00000 | Uiso | 1.00 |
| H54 | H | 0.43438 | 0.67496 | 0.13879 | 0.00000 | Uiso | 1.00 |
| H55 | H | 0.46301 | 0.78204 | 0.22402 | 0.00000 | Uiso | 1.00 |
| H56 | H | 0.51767 | 0.88032 | 0.14098 | 0.00000 | Uiso | 1.00 |
| H57 | H | 0.51530 | 0.78744 | 0.08665 | 0.00000 | Uiso | 1.00 |
| H58 | H | 0.73769 | 0.83551 | 0.24300 | 0.00000 | Uiso | 1.00 |
| H59 | H | 0.69524 | 0.70472 | 0.25101 | 0.00000 | Uiso | 1.00 |
| H60 | H | 0.69072 | 0.81080 | 0.32069 | 0.00000 | Uiso | 1.00 |
| H61 | H | 0.98211 | 1.02676 | 0.62065 | 0.00000 | Uiso | 1.00 |
| H62 | H | 1.15010 | 0.96615 | 0.72590 | 0.00000 | Uiso | 1.00 |
| H63 | H | 1.06616 | 1.01664 | 0.76417 | 0.00000 | Uiso | 1.00 |
| H64 | H | 1.03969 | 0.84687 | 0.74802 | 0.00000 | Uiso | 1.00 |
| H65 | H | 1.00510 | 0.83410 | 0.64268 | 0.00000 | Uiso | 1.00 |
| H66 | H | 0.79154 | 0.90640 | 0.53074 | 0.00000 | Uiso | 1.00 |
| H67 | H | 0.83251 | 0.93665 | 0.65811 | 0.00000 | Uiso | 1.00 |
| H68 | H | 0.83316 | 0.76230 | 0.68178 | 0.00000 | Uiso | 1.00 |
| H69 | H | 0.81562 | 0.73071 | 0.58623 | 0.00000 | Uiso | 1.00 |
| H70 | H | 0.61644 | 0.70373 | 0.54393 | 0.00000 | Uiso | 1.00 |
| H71 | H | 0.60103 | 0.79436 | 0.45185 | 0.00000 | Uiso | 1.00 |
| H72 | H | 0.61714 | 0.84255 | 0.56348 | 0.00000 | Uiso | 1.00 |
| H73 | H | 0.23237 | 1.01081 | 0.84819 | 0.00000 | Uiso | 1.00 |
| H74 | H | 0.37678 | 0.98260 | 0.95540 | 0.00000 | Uiso | 1.00 |
| H75 | H | 0.33389 | 1.09739 | 0.93189 | 0.00000 | Uiso | 1.00 |
| H76 | H | 0.27302 | 1.02156 | 0.95924 | 0.00000 | Uiso | 1.00 |
| H77 | H | 0.92845 | 0.34726 | 0.29651 | 0.00000 | Uiso | 1.00 |
| H78 | H | 0.85564 | 0.34756 | 0.17184 | 0.00000 | Uiso | 1.00 |
| H79 | H | 0.83467 | 0.43444 | 0.22113 | 0.00000 | Uiso | 1.00 |
| H80 | H | 0.97006 | 0.42799 | 0.25162 | 0.00000 | Uiso | 1.00 |
| H81 | H | 0.40974 | -0.08366 | 0.57357 | 0.00000 | Uiso | 1.00 |
| H82 | H | 0.40191 | -0.05445 | 0.48091 | 0.00000 | Uiso | 1.00 |
| H83 | H | 0.32424 | 0.01191 | 0.49763 | 0.00000 | Uiso | 1.00 |
| H84 | H | 0.47946 | 0.01857 | 0.58567 | 0.00000 | Uiso | 1.00 |
| H85 | H | 0.17156 | -0.02595 | 0.56907 | 0.00000 | Uiso | 1.00 |
| H86 | H | 0.08359 | 0.07275 | 0.53422 | 0.00000 | Uiso | 1.00 |
| H87 | H | 0.14981 | 0.01190 | 0.64117 | 0.00000 | Uiso | 1.00 |
| H88 | H | 0.23755 | 0.07955 | 0.63623 | 0.00000 | Uiso | 1.00 |

(NH$_3$BH$_2$NH$_2$BH$_2$NH$_2$BH$_3$), unit cell optimised
```
data_32_500eV\fine+cell
_audit_creation_date          2022-04-05
_audit_creation_method        'Materials Studio'
_symmetry_space_group_name_H-M   'P21/C'
_symmetry_Int_Tables_number   14
_symmetry_cell_setting        monoclinic
loop_
_symmetry_equiv_pos_as_xyz
 x,y,z
 -x,y+1/2,-z+1/2
 -x,-y,-z
 x,-y+1/2,z+1/2
_cell_length_a                14.0521
_cell_length_b                12.5784
_cell_length_c                19.2528
_cell_angle_alpha             90.0000
_cell_angle_beta              128.7292
_cell_angle_gamma             90.0000
loop_
_atom_site_label
_atom_site_type_symbol
_atom_site_fract_x
_atom_site_fract_y
_atom_site_fract_z
_atom_site_U_iso_or_equiv
_atom_site_adp_type
_atom_site_occupancy
B1    B   0.61234  0.14161  0.69593  0.05000  Uiso  1.00
N1    N   0.74985  0.13798  0.78118  0.05000  Uiso  1.00
B2    B   0.80335  0.24827  0.83148  0.05000  Uiso  1.00
N2    N   0.94601  0.24008  0.90513  0.05000  Uiso  1.00
B3    B   1.00612  0.33672  0.97390  0.05000  Uiso  1.00
H1    H   0.54946  0.17792  0.71315  0.05000  Uiso  1.00
H2    H   0.60187  0.18975  0.63663  0.05000  Uiso  1.00
H4    H   0.76311  0.08260  0.82641  0.05000  Uiso  1.00
H5    H   0.80196  0.11174  0.76378  0.05000  Uiso  1.00
H6    H   0.78333  0.31438  0.77695  0.05000  Uiso  1.00
H7    H   0.75652  0.27053  0.86549  0.05000  Uiso  1.00
H8    H   0.98548  0.23333  0.87482  0.05000  Uiso  1.00
H9    H   0.97069  0.17067  0.94111  0.05000  Uiso  1.00
H10   H   1.00173  0.31692  1.03399  0.05000  Uiso  1.00
H11   H   0.95188  0.41887  0.93541  0.05000  Uiso  1.00
H12   H   1.11486  0.34492  1.00700  0.05000  Uiso  1.00
B4    B   0.50969  0.43569  0.71360  0.05000  Uiso  1.00
N3    N   0.46702  0.38819  0.62292  0.05000  Uiso  1.00
B5    B   0.33020  0.34237  0.55254  0.05000  Uiso  1.00
N4    N   0.25702  0.34528  0.59044  0.05000  Uiso  1.00
B6    B   0.17134  0.44485  0.56525  0.05000  Uiso  1.00
H13   H   0.50674  0.36814  0.75773  0.05000  Uiso  1.00
H15   H   0.45024  0.51405  0.70061  0.05000  Uiso  1.00
H16   H   0.47623  0.44544  0.58905  0.05000  Uiso  1.00
H17   H   0.52384  0.32667  0.63483  0.05000  Uiso  1.00
H18   H   0.33803  0.24984  0.53711  0.05000  Uiso  1.00
H19   H   0.27691  0.39738  0.48626  0.05000  Uiso  1.00
H20   H   0.31598  0.33436  0.65830  0.05000  Uiso  1.00
H21   H   0.20393  0.27761  0.56705  0.05000  Uiso  1.00
H22   H   0.22991  0.52690  0.58489  0.05000  Uiso  1.00
H23   H   0.12913  0.44167  0.60495  0.05000  Uiso  1.00
```

```
H24   H   0.08772   0.44511   0.48540   0.05000  Uiso  1.00
B7    B   0.27345   0.57022  -0.00885   0.05000  Uiso  1.00
N5    N   0.26635   0.66818  -0.06284   0.05000  Uiso  1.00
B8    B   0.22211   0.77997  -0.05360   0.05000  Uiso  1.00
N6    N   0.33300   0.83259   0.03608   0.05000  Uiso  1.00
B9    B   0.44363   0.86757   0.03827   0.05000  Uiso  1.00
H25   H   0.31804   0.49323  -0.01834   0.05000  Uiso  1.00
H26   H   0.33068   0.59327   0.06945   0.05000  Uiso  1.00
H28   H   0.34950   0.68144  -0.04834   0.05000  Uiso  1.00
H29   H   0.20852   0.64938  -0.12983   0.05000  Uiso  1.00
H30   H   0.19464   0.83592  -0.11562   0.05000  Uiso  1.00
H31   H   0.13695   0.76681  -0.05285   0.05000  Uiso  1.00
H32   H   0.36163   0.78334   0.08860   0.05000  Uiso  1.00
H33   H   0.29991   0.89825   0.04691   0.05000  Uiso  1.00
H34   H   0.41205   0.94339  -0.01110   0.05000  Uiso  1.00
H35   H   0.53224   0.88970   0.11504   0.05000  Uiso  1.00
H36   H   0.47107   0.79357   0.01180   0.05000  Uiso  1.00
B10   B   1.01084   1.09682   0.78046   0.05000  Uiso  1.00
N7    N   1.00571   0.98490   0.81574   0.05000  Uiso  1.00
B11   B   0.89496   0.91200   0.74411   0.05000  Uiso  1.00
N8    N   0.88698   0.81275   0.79116   0.05000  Uiso  1.00
B12   B   0.76080   0.75282   0.73529   0.05000  Uiso  1.00
H37   H   0.91116   1.14046   0.73976   0.05000  Uiso  1.00
H38   H   1.08492   1.15377   0.84346   0.05000  Uiso  1.00
H39   H   1.03831   1.08470   0.73181   0.05000  Uiso  1.00
H40   H   1.00795   0.99774   0.86972   0.05000  Uiso  1.00
H41   H   1.08598   0.94429   0.84318   0.05000  Uiso  1.00
H42   H   0.90823   0.88165   0.69059   0.05000  Uiso  1.00
H43   H   0.79993   0.96281   0.70640   0.05000  Uiso  1.00
H44   H   0.95736   0.76075   0.81376   0.05000  Uiso  1.00
H45   H   0.90141   0.83918   0.84788   0.05000  Uiso  1.00
H46   H   0.68411   0.81054   0.72636   0.05000  Uiso  1.00
H48   H   0.76990   0.66919   0.77091   0.05000  Uiso  1.00
N9    N   0.64892   0.47383   0.77073   0.05000  Uiso  1.00
H50   H   0.65955   0.53699   0.74190   0.05000  Uiso  1.00
H51   H   0.70708   0.41277   0.78183   0.05000  Uiso  1.00
H52   H   0.68082   0.49842   0.83280   0.05000  Uiso  1.00
N10   N   0.13759   0.54192  -0.04916   0.05000  Uiso  1.00
H53   H   0.10005   0.59620  -0.03170   0.05000  Uiso  1.00
H54   H   0.12833   0.46828  -0.02971   0.05000  Uiso  1.00
H55   H   0.07927   0.54020  -0.11808   0.05000  Uiso  1.00
N11   N   0.56770   0.02149   0.66238   0.05000  Uiso  1.00
H57   H   0.48257   0.01895   0.60065   0.05000  Uiso  1.00
H58   H   0.62717  -0.02152   0.65896   0.05000  Uiso  1.00
H59   H   0.55927  -0.01953   0.70490   0.05000  Uiso  1.00
N12   N   0.27916   0.22761   0.86024   0.05000  Uiso  1.00
H61   H   0.36599   0.19486   0.90001   0.05000  Uiso  1.00
H63   H   0.28219   0.29363   0.89304   0.05000  Uiso  1.00
H64   H   0.22046   0.17569   0.85841   0.05000  Uiso  1.00
```

## 13. Crystal structure (VBH)[B(C$_6$H$_5$)$_4$]

**Table 13.1.** Crystal structure parameters of (C$_{18}$H$_{39}$N$_4$PH)[B(C$_6$H$_5$)$_4$].

| Compound | (C$_{18}$H$_{39}$N$_4$PH)[B(C$_6$H$_5$)$_4$] |
|---|---|
| K$_\alpha$(Å) | 1.54184 (Cu) |
| Temperature (K) | 100(2) |
| Space group | $P\bar{1}$ |
| Z | 4 |
| a (Å) | 11.7376(3) |
| b (Å) | 19.5388(5) |
| c (Å) | 20.5479(4) |
| α (°) | 61.751(2) |
| β (°) | 73.618(2) |
| γ (°) | 89.605(2) |
| V (Å$^3$) | 3937.71(18) |
| $\rho_{calc.}$ (g cm$^{-3}$) | 1.118 |
| $\mu_{exp.}$ (mm$^{-1}$) | 0.856 |
| $\vartheta_{max}$ (°) | 75.2030 |
| $R_1$ | 0.0695 |
| $wR_2$ | 0.2094 |
| GooF | 1.048 |
| Crystal size (mm×mm×mm) | 0.06 x 0.16 x 0.20 |
| Crystal colour | colorless |
| CCDC No. | |

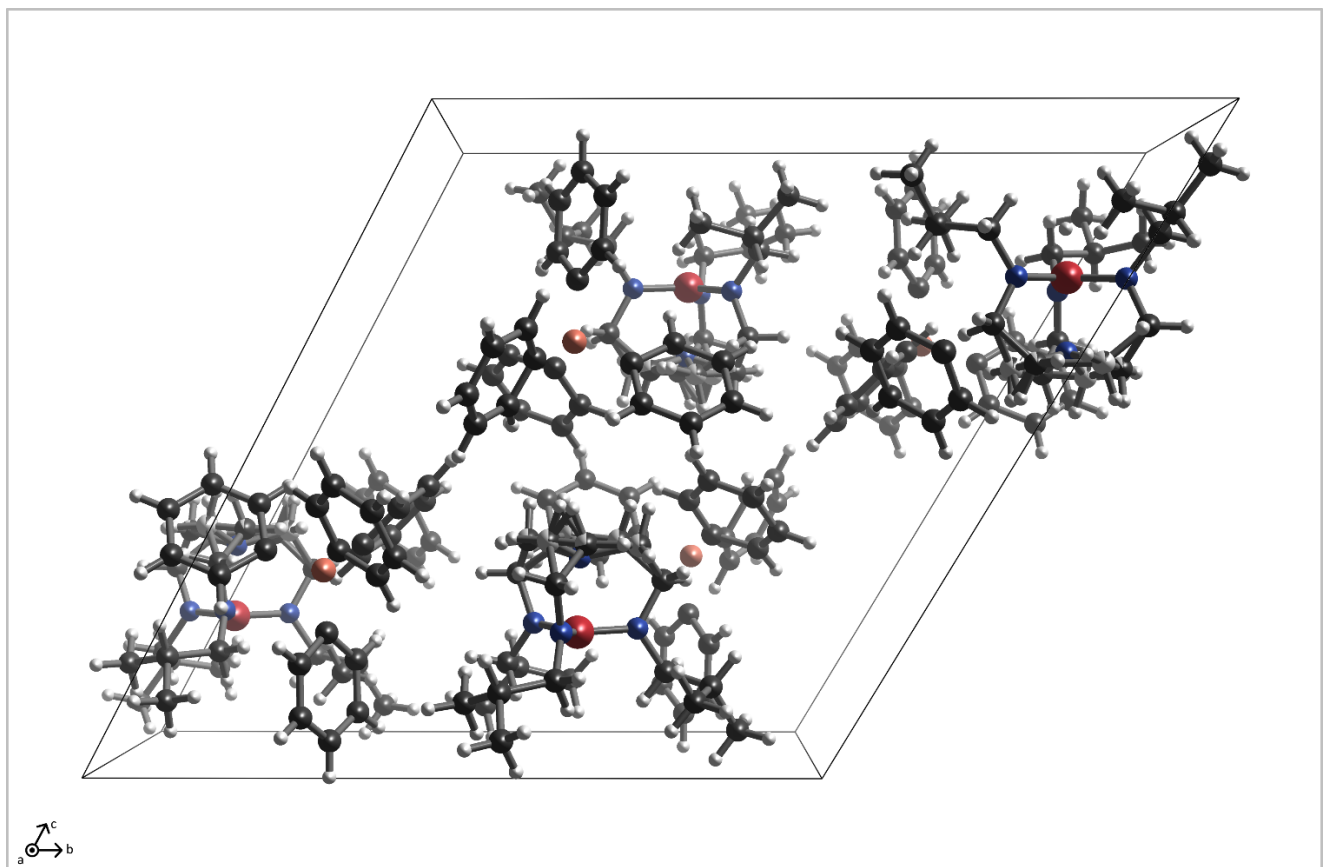

**Fig. S13.2.** Visualisation of the unit cell of the side product: (C$_{18}$H$_{39}$N$_4$PH)[B(C$_6$H$_5$)$_4$].